\documentclass[12pt,letterpaper]{article}
\pdfoutput=1
\usepackage{ifpdf}
\usepackage{subcaption}
\usepackage{jheppub}	
\usepackage{epsfig}
\usepackage{enumitem}
\usepackage[latin1]{inputenc}
\usepackage{bbm,amsfonts}
\usepackage{graphicx}
\usepackage{amssymb,amsmath,amsfonts,mathtools}
\usepackage{fancybox}
\usepackage{enumerate}
\usepackage{wrapfig}
\usepackage{dsfont}
\usepackage{verbatim}
\usepackage{tikz}
\usepackage{comment}
\usepackage{wrapfig}
\usepackage{slashed}
\usepackage{shuffle}
\usepackage{subcaption}
\usepackage{braket}

\author[a]{Sara Bonansea,}
\author[b]{Silvia Davoli,}
\author[b]{Luca Griguolo,}
\author[a]{and Domenico Seminara}
\affiliation[a]{Dipartimento di Fisica, Universit\`a di Firenze and INFN Sezione di Firenze, via G. Sansone 1, 50019 Sesto Fiorentino, Italy}
\affiliation[b]{Dipartimento SMFI, Universit\`a di Parma and INFN Gruppo Collegato di Parma, Viale G.P. Usberti 7/A, 43100 Parma, Italy\\}

\emailAdd{sara.bonansea@fi.infn.it}
\emailAdd{silvia.davoli@fis.unipr.it} 
\emailAdd{luca.griguolo@fis.unipr.it} 
\emailAdd{seminara@fi.infn.it}

\abstract{We consider circular Wilson loops in a defect version of $\mathcal{N}=4$ super-Yang-Mills theory which is dual to the D3-D5 brane system with $k$ units of flux. When the loops are parallel to the defect, we can construct both BPS and non-BPS operators, depending on the orientation of the scalar couplings in the R-symmetry directions. At strong 't Hooft coupling we observe, in the non supersymmetric case, a Gross-Ooguri-like phase transition in the dual gravitational theory: the familiar disk solution dominates, as expected, when the operator is far from the defect while a cylindrical string worldsheet, connecting the boundary loop with the probe D5-brane, is favourite below a certain distance (or equivalently for large radii of the circles). In the BPS case, instead, the cylindrical solution does not exist for any choice of the physical parameters, suggesting that the exchange of light supergravity modes always saturate the expectation value at strong coupling. We study the double-scaling limit for large $k$ and large 't Hooft coupling, finding full consistency in the non-BPS case between the string solution and the one-loop perturbative result. Finally we discuss, in the BPS case, the failure of the double-scaling limit and the OPE expansion of the Wilson loop, finding consistency with the known results for the one-point functions of scalar composite operators.  }

\title{Circular Wilson loops in defect ${\cal N}$=4 SYM: phase transitions, double-scaling limits and OPE expansions}

\newcommand{\be}{\begin{equation}}
\newcommand{\ee}{\end{equation}}
\newcommand{\beq}{\begin{equation}}
\newcommand{\eeq}{\end{equation}}
\newcommand{\bea}{\begin{eqnarray}}
\newcommand{\eea}{\end{eqnarray}}
\newcommand{\ena}{\end{eqnarray}}

\newcommand{\p}{\pi}

\numberwithin{equation}{section}

\def\clock{{\count0=\time
           \divide\count0 60
           \ifnum\count0<10 0\fi\the\count0
           \multiply\count0 -60 \advance\count0 \time
           :\ifnum\count0<10 0\fi \the\count0
         }}
\newcommand{\timestamp}{{\small\vbox{\hbox{\tt\jobname.tex}
\hbox{\the\day/\the\month/\the\year, \clock}}}}
\usepackage{transparent}
\begin{document}

\maketitle
\allowdisplaybreaks

\section{Introduction}
The well-established paradigm of AdS/CFT opened the possibility to explore, al least at large-N, the strong coupling regime of four-dimensional gauge theories, obtaining results that have been
confirmed through the application of non-perturbative techniques, as duality, localization, integrability and bootstrap. These methods produce, in principle, answers that interpolate between
weak and strong coupling allowing a precise comparison with the gauge-gravity predictions. Unfortunately, many properties rely heavily on large amounts of supersymmetry or, even more crucially,
on conformal symmetry, making difficult the application to the real world. Any attempt to extend the validity of these approaches to less symmetric situations is certainly welcome. A quite general possibility to reduce the amount of symmetry in quantum field theory is
to introduce a defect or an interface into the game: starting from some (super)conformal theory
we can introduce, for example, a domain-wall preserving a subset of the original invariance. In this case, one generally obtains a defect Conformal Field Theory (dCFT), in which new degrees
of freedom living on the defect interact non-trivially with the bulk. Of particular interest are dCFTs with holographic duals.   A certain number of examples of this type exists, following the original
idea presented in \cite{Karch:2001cw,Karch:2000gx,DeWolfe:2001pq,Erdmenger:2002ex}. In this paper, we will consider ${\cal N}=4$ supersymmetric Yang-Mills theory (${\cal N} = 4$ SYM theory) with a codimension-one defect located at $x_3 = 0$: it separates two regions of space-time where the gauge group is respectively $SU(N)$ and $SU(N-k)$ \cite{Nagasaki:2011ue}. In the field theory description, the difference in the rank of the gauge group is related to a non-vanishing vacuum expectation value (VEV) proportional to $1/x_3$, assigned to three of the ${\cal N}=4$ SYM scalar fields in the region $x_3 > 0$. The VEV originates from the boundary conditions on the defect that are chosen to preserve part of the original supersymmetry.  On the other hand, the gauge theory is dual to a D5-D3 probe-brane system involving a single D5 brane whose profile spans $AdS_4\times S^2$, in the presence of a background flux of $k$ units through the $S^2$. The flux $k$ controls the VEV of the scalar fields and represents a new tunable parameter in the usual ${\cal N}=4$ SYM framework, which can be used to probe the theory in different regimes. 
In the last few years there has been a certain amount of work in studying such a system: in particular the vacuum expectation value for a large class of scalar operators has been obtained, both at weak coupling \cite{Buhl-Mortensen:2016pxs}, using perturbation theory, and at strong coupling, by means of the dual-brane set-up \cite{Nagasaki:2012re,Kristjansen:2012tn,Buhl-Mortensen:2016jqo}. A particular feature of dCFT is that one-point functions can be different from zero, and this fact has been largely exploited in these investigations. More recently, a serious attempt to extend the integrability program in this context has been performed by the NBI group \cite{deLeeuw:2015hxa,Buhl-Mortensen:2015gfd,Buhl-Mortensen:2017ind}, leading to some interesting generalizations of the original techniques.   

Moreover, the presence of the extra-parameter $k$ allows for a new kind of double-scaling limit, able to connect, in principle, the perturbative regime with the gauge-gravity computations. It consists of sending the 't Hooft coupling $\lambda$ as well as $k^2$ to infinity while keeping fixed the ratio of the two parameters: the perturbative expansion organizes in powers of this ratio, that can be considered small.  At the same time, the large 't Hooft coupling still supports the validity of the dual gravity calculations. Thus, in that regime, one could try to successfully compare gauge and gravity results, providing a new non-trivial verification of the AdS/CFT correspondence \cite{Buhl-Mortensen:2016jqo}. One-point functions of local operators, both at tree-level and one-loop, match the AdS/CFT predictions accurately in the double-scaling limit.
Further studies on the two-point functions, OPE and boundary OPE has been recently performed in \cite{deLeeuw:2017dkd}. Less attention has been instead devoted to other natural observables that AdS/CFT correspondence can explore in this context, namely  Wilson loops.   At strong coupling, the vacuum expectation value of these operators are computed by evaluating the area
of the minimal surface spanned by the fundamental string in the supergravity dual, with boundary conditions dictated by the contour and the scalar couplings \cite{Rey:1998ik,Maldacena:1998im}. Their supersymmetric version \cite{Zarembo:2016bbk} can be often evaluated exactly through localization techniques, allowing a precise interpolation between weak and strong coupling \cite{Erickson:2000af,Drukker:2000rr,Pestun:2007rz}. In the presence of defects, Wilson loop operators were first considered in \cite{Nagasaki:2011ue}: their expectation values have been studied in the double-scaling limit, allowing to compare perturbation theory successfully to the string calculation in the case of quark-antiquark potential \cite{Nagasaki:2011ue,deLeeuw:2016vgp}. More recently, circular Wilson loops, analog to the supersymmetric ones in ordinary ${\cal N }= 4$ super Yang-Mills, have been examined in \cite{Aguilera-Damia:2016bqv}, producing some interesting results. There it was considered a circular Wilson loop of radius $R$ placed at distance $L$ from the defect and parallel to it, whose internal space orientation has been parameterized by an angle $\chi$. Its vacuum expectation value has been computed both at weak and strong coupling, and, in the double-scaling limit and for small $\chi$ and small $L/R$, the results appeared consistent. 

In this paper, we investigate further the same circular Wilson loop in defect ${\cal N} = 4$ super Yang-Mills theory, generalizing the computations presented in \cite{Aguilera-Damia:2016bqv} both at strong and weak-coupling. In particular we are able to cover the full parameter space of the string solution of our system: we derive the exact solution for the minimal surface, describing the Wilson loop in the AdS/CFT setting, for any value of the flux $k$, angle $\chi$ and ratio $L/R$ and we can explore its complicated structure in different regions of the parameters.  
\begin{figure}[b]   
\centering      
	\includegraphics[width=.78\textwidth]{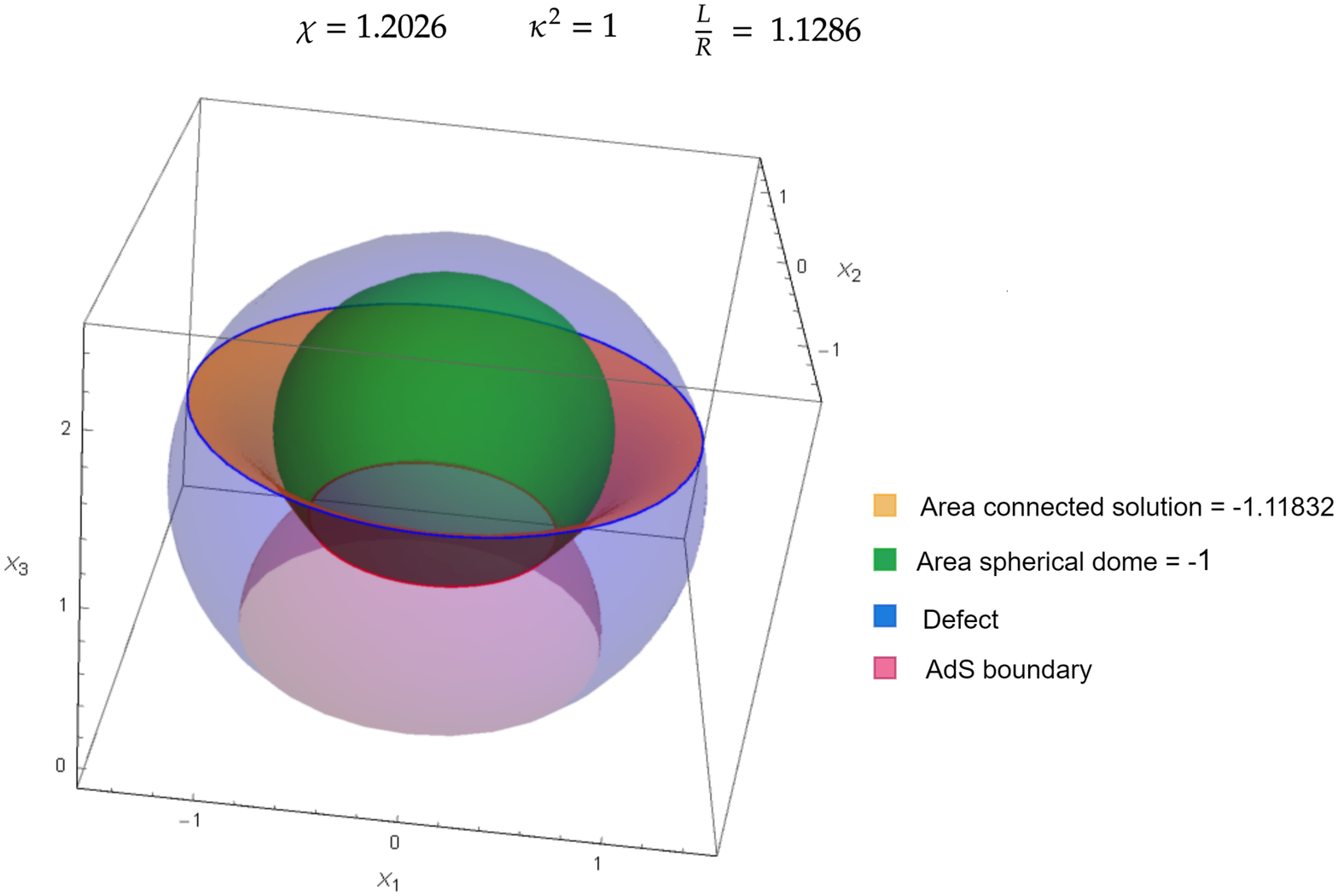}
		\caption{\label{fig.6vbn} \footnotesize  The projection of the minimal surface in AdS$_5$ is entirely contained in a sphere $S^3$ (see app. \ref{circles}). In this plot this sphere is mapped into $\mathds{R}^3$ through the usual stereographic projection. The blue  spherical cap is the intersection between $S^3$ and the D5 while pink one is the intersection between $S^3$ and the boundary of AdS$_5$.  Then the green dome  and the  yellow surface connecting the boundary with the   D5 are the two competing solutions. For this choice of parameters the dominant solution is the connected one. }
\end{figure}Nicely we recover, in the limit of large $k$, the result of \cite{Aguilera-Damia:2016bqv} without restrictions on $L/R$ and $\chi$. The main output of our analysis is the discovery of a first-order phase transition of Gross-Ooguri type: for any flux $k$ and any non-zero angle $\chi$ the disk solution (describing the Wilson loop in the absence of defect) still exists and dominates, as expected, when the operator is far from the defect. On the other hand, our cylindrical string solution, connecting the boundary loop with the probe D5-brane, is favorite below a certain distance (or equivalently for a large radius of the circles). We can compare the classical actions associated with the solutions, by a mixture of analytical and numerical methods, finding the critical ratio $L/R$ as a function of $k$ and $\chi$. A related investigation has been performed in \cite{Preti:2017fhw} for the quark-antiquark potential. A second important conclusion is that in the BPS case, that corresponds to $\chi=0$, the cylindrical solution does not exist for any choice of the physical parameters, suggesting that exchanges of light supergravity modes always saturate the expectation value at strong coupling. This behavior strongly resembles an analog result for correlators of relatively BPS Wilson loops in ${\cal N}=4$ SYM \cite{Drukker:2007qr}, which can also be exactly computed through localization \cite{Giombi:2009ms,Bassetto:2009rt}. The weak coupling analysis corroborates the exceptionality of the BPS case: the first non-trivial perturbative contribution is evaluated exactly in terms of a Mejer-G function, and its large $k$ expansion does not scale in a way to match the string solution. In particular, it is not possible to recover the large $k$ limit starting from the equivalent asymptotic expansion of the $\chi\neq 0$ case: the order of limits does not commute. In the  regime $L/R\to \infty$, we expect instead that the perturbative result could be understood in term of the OPE expansion of the Wilson loop: we confirm this idea, and we reconstruct the first two non-trivial terms of the expansion from the known results for the one-point function of scalar operators.

The structure of the paper is the following. In sec. \ref{Prel} we start with discussing the generalities of the problem while in sec. \ref{section:classicsolution} and sec. \ref{subsection:2.1} we present the general solution of the equation of motions: we obtain an explicit expression for the functions $y(\sigma), r(\sigma), x_3(\sigma),\theta(\sigma)$ that describe the embedding of the string worldsheet into $AdS_5\times S^5$ ($\sigma$ is the spatial worldsheet parameter, see eqs. \eqref{domea}. and \eqref{ansatza})
\be
\begin{gathered}
y(\sigma)=\frac{ R\cosh\eta}{\sqrt{1+g^2(\sigma)}} \textrm{sech} [v(\sigma)-\eta]\quad\quad
r(\sigma)=R\cosh\eta \frac{g(\sigma)}{\sqrt{1+g^2(\sigma)}} \textrm{sech} [v(\sigma)-\eta]\\
x_3(\sigma)= -R\cosh\eta\tanh [v(\sigma)-\eta]\quad\quad \theta(\sigma)=j\sigma+\chi.
\end{gathered}
\ee
where the function $v(\sigma)$ is defined by
 \be
 \label{eq:2.21aab}
 v'(\sigma)=\sqrt{-\frac{\left(j^2+m\right) \left(j^2 m+1\right)}{(m+1)^2}}\frac{1}{1+g^2(\sigma)},
\ee
and 
\be
\label{eq:2.43b}
\eta = v(\tilde\sigma)+\tanh ^{-1}\left(-\sqrt{-\frac{(m+1)^2}{\left(j^2+m\right) \left(j^2 m+1\right)}}
\frac{g^\prime(\tilde\sigma )}{g(\tilde\sigma
  )}\right).
\ee
The function $g(\sigma)$, that controls the full construction, has the explicit expression
\be
\label{bobob}
g(\sigma)=\sqrt{\frac{j^2-1}{m+1}}\text{ns}\left(\sqrt{\frac{j^2-1}{m+1}}\sigma, m\right).
\ee
The minimal surface is given in terms of three real constants $(m,j,\tilde\sigma)$ ($\tilde\sigma$ is the limiting value of the worldsheet coordinate) that are highly nonlinearly related to the physical parameters ($L/R,\chi, \kappa$) once the boundary conditions are imposed: we have defined $\kappa =\frac{\pi k}{\sqrt{\lambda}}$. Sec. \ref{allowedparameters} is devoted to finding the parameter space of the string solution, using the appropriate boundary conditions and some positivity requirements: the analysis can be performed restricting this moduli space into two regions that we call A and B (see equations (\ref{eq:2.51a})). In the limiting case $\chi=0$, both regions shrink to zero. Sec. \ref{section:structure} is the heart of our investigations, in which we discuss the structure of the connected string solution: the existence of the minimal surface is discussed as a function of the ratio $L/R$, and we find that there is a limiting value beyond that the solution ceases to exist. Moreover, there exist regions where a second branch appears, showing the presence of two competing connected solutions. For the sake of clarity, we display here the final result of this analysis, that singles out a critical angle, $\chi_s$, distinguishing two situations.

\paragraph{\bf(1)} $\boldsymbol{\chi_s\leq\chi\le\frac{\pi}{2}}$: In this case we have always two branches for the solution, no matter of the value of the flux $\kappa$

\paragraph{\bf(2)} $\boldsymbol{0<\chi<\chi_s}$: In this region we can determine a critical value of the flux $\kappa _s(\chi)$: above this value we have a single branch solution while below a second branch appears.

The evaluation of the area of the minimal surface, obtained by computing the Polyakov action on the solutions is done in sec. \ref{subsect:3.2}. We find that the dominant branch is always physically connected to a vanishing distance from the defect. Finally in section \ref{subsect:3.3} we compare the area of the dominant connected solution with the disk-like on-shell action: in the different regions we always find that decreasing $L/R$ from $+\infty$, where obviously the disk-like surface is the relevant saddle-point, there exists a critical value (depending on $(\chi,k)$) below which the connected cylinder starts being dominant. Nicely the disk solution (the spherical $dome$ as we will call it later) ceases to be dominant before touching the D5 brane profile. The last section is devoted to some perturbative computations, checking the picture emerging from strong coupling: first in sect. \ref{NBPS} we briefly recall the computation of the non-BPS Wilson loop at the first two perturbative orders and perform its double-scaling limit. Then  in sect. \ref{subsect:strongcouplingexpansion} we explicitly expand in $\lambda/k^2$ the AdS/CFT solution. We recover the result of \cite{Aguilera-Damia:2016bqv} without compromising ourselves with the value of other parameters (taking $L/R$ and $\chi$ generic) and showing the consistency with the relevant order of weak-coupling perturbation theory. Then we discuss the case $\chi=0$ in
 sec.\ref{perturbBPS}, remarking its peculiarity and highlighting the absence of a string counterpart. Finally, after having review the standard OPE expansion for the circular Wilson loop in sec. \ref{OPE1},  we discuss the OPE picture of the present BPS case in sec. \ref{OPE}, finding consistency of our results with the known computations of some scalar one-point functions. Our conclusions and a list of interesting future follow-ups of our investigations are presented in sec. \ref{conclusions}. A certain number of technical appendices complete the paper.

\section{Prelude}
\label{Prel}
The goal of the present paper is to study the vacuum expectation value of a circular Maldacena-Wilson loop in a four-dimensional dCFT given by 
$\mathcal{N}=4$ SYM theory with a co-dimension one hyperplane inserted at $x_{3}=0$ as in \cite{DeWolfe:2001pq,Nagasaki:2011ue,Buhl-Mortensen:2016jqo}. 
More precisely, the defect separates two different $\mathcal{N}=4$ SYM theories: in the region $x_{3}<0$, we have the standard $\mathcal{N}=4$
SYM with gauge group $SU\left(N-k\right)$. On the other hand, an Higgsed $\mathcal{N}=4$ SYM lives in the $x_{3}>0$ region, with gauge group $SU\left(N\right)$,
where three scalar fields receive a $x_{3}$-dependent VEV. At the level of the field theory, the picture is the following.
The action for the dCFT is composed by two terms
\begin{equation}
S=S_{\mathcal{N}=4}+S_{D=3},
\end{equation}
where $S_{\mathcal{N}=4}$ is the usual $\mathcal{N}=4$ SYM action
that describes the bulk of the space-time, while $S_{D=3}$ accounts
in general for degrees of freedom sited on the defect: they could both self-interact and couple to the bulk $\mathcal{N}=4$ SYM. The presence of the defect implies that fields
living in the $x_{3}>0$ region will have a non-trivial vacuum solution: by imposing that  a part of the supersymmetry is preserved a specific profile is obtained for the scalars.
Following \cite{Nagasaki:2011ue}, one assumes  the ansatz
\begin{equation}
A_{\mu}=0\,\left(\mu=0,1,2,3\right),\qquad\Phi_{I}=\Phi_{I}(x_3) \,\left(I,J,K=1,2,3\right),\qquad\Phi_{M}=0\,\left(M=4,5,6\right).
\end{equation}
and the vanishing of fermions supersymmetry variation
\begin{equation}
0=\delta\psi=\partial_{3}\Phi_{I}\tilde\Gamma^{3I}\epsilon-\frac{i}{2}\left[\Phi_{I},\Phi_{J}\right]\tilde\Gamma^{IJ}\epsilon,
\end{equation}
leads to the Nahm\textquoteright s equations:
\begin{equation}
\partial_{3}\Phi_{I}=-\frac{i}{2}\epsilon_{IJK}\left[\Phi_{I},\Phi_{K}\right],\label{eq:dCFT Nahm's equations}
\end{equation}
with $\epsilon$ satisfying
\begin{equation}
\left(1-\Gamma^{3456}\right)\epsilon=0.
\end{equation}
We have followed the notation of \cite{Buhl-Mortensen:2016jqo} and introduced $\tilde \Gamma^I=\Gamma^{I+3}$,  $\tilde \Gamma^{3I}=\Gamma^3\Gamma^{I+3}$ and $\tilde \Gamma^{IJ}=\Gamma^{I+3}\Gamma^{J+3}$.
The solution to eq. (\ref{eq:dCFT Nahm's equations}) it is known
\cite{Constable:1999ac} and it is called ``fuzzy funnel'' solution,
reading 
\begin{equation}
\left\langle \Phi_{I}\right\rangle _{\textrm{tree}}=\Phi_{I}^{\textrm{cl}}=-\frac{1}{x_{3}}t_{I}\oplus0_{\left(N-k\right)\times\left(N-k\right)},\label{eq:Fuzzy funnel VEV}
\end{equation}
where $t_{I}$ are generators of a representation of $SU\left(2\right)$
(we can choose, for example, $I=1,2,3$). This means that the $t_{I}$
are $k\times k$ matrices satisfying
\begin{equation}
\left[t_{I},t_{J}\right]=i\epsilon_{IJK}t_{K},\qquad I,J,K=1,2,3.
\end{equation}
All the other classical fields are zero. We observe that the $SO(6)$ R-symmetry of the original $\mathcal{N}=4$ SYM action is reduced to $SO(3)\times SO(3)$. We would like to study the expectation value of circular Maldacena-Wilson loops in this vacuum. A natural choice is to center the circle along the $x_3$ axis at a distance $L$ from the defect, i.e. $C=(0,0,0,L)$. The radius of the circle is  $R$ and it extends only along the transverse  directions
 $x_1$ and $x_2$, namely
 \be
 x^\mu(\tau)=(0,R\cos\tau,R\sin\tau,L)
 \ee
The residual $SO(3)\times SO(3)$ symmetry suggests to couple only two scalars to the Wilson loop: one massless $\Phi_6$ and one massive  $\Phi_3$ and we get
\be
\mathcal{W}=\mathrm{Pexp}\left(i\oint (A_\mu\dot x^\mu+i |\dot x|(\Phi_3\sin\chi+\Phi_6\cos\chi))\right),
\ee
where the angle $\chi$ parametrizes the  strength of the coupling with the two scalars.  Because of the conformal invariance $\langle \mathcal{W}\rangle$ does not depend separately   by $L$ and $R$ but only through the ratio $R/L$. Moreover, the explicit analysis performed in \cite{Aguilera-Damia:2016bqv} shows that in the absence of the defect our observable is
always 1/2 BPS, but  in its presence all the supercharges are broken unless $\chi=0.$

\subsection{Setting-up the geometric  description}
\label{section:classicsolution}
On  the string theory side, the field theory picture  translates into a system of  $N$ $D3-$branes  intersecting a single $D5-$brane,
where $k$  $D3-$branes out of the stack of $N$ terminate on it. In the near horizon limit  we can
view the  D5 as a probe brane\footnote{This picture of probe $D5-$branes
in $AdS_5\times S^5$ holds when the number $M$ of D5 is much less than $N$.} moving in $AdS^5\times S^5$. 
The intersection between D3 and D5 mimics  the presence of a defect (domain wall) of codimension
one located at $x_3=0$ in the field theory. 
The $AdS_5$ is  is parametrized  in Poincar\`e  coordinates  where the metric takes the form
\be
ds^2_{AdS_5}=\frac{1}{y^2}(-d t^2 +dy^2+dr^2+r^2 d\phi^2+dx_3^2).
\ee
and for the sphere $S^5$   we write 
\be
ds^2=d\theta^2+\sin^2 \theta d \Omega^2_{(1)}+\cos^2\theta  d\Omega^2_{(2)},
\ee
where $d\Omega^2_{(i)}=d\alpha_i^2+\sin^2\alpha_id\beta_i^2$ denotes the metric of  the two $S^2$ inside the $S_5$. In these coordinates
the $D5-$brane solution wraps the sphere $\Omega_{(1)}$ and its embedding in the target space is given by \cite{Nagasaki:2011ue}
\be
y=\frac{1}{\kappa} x_3, \quad\quad \theta=\frac{\pi}{2},\quad\quad \alpha_2=\alpha^{(0)}_2\quad\quad\mathrm{and}
\quad\quad\beta_2=\beta^{(0)}_2,
\ee
where $\alpha_2^{(0)}$ and $\beta_2^{(0)}$ are two constant values.  There is also an abelian background gauge field providing a non-trivial flux through  $\Omega_{(1)}$, {\it i.e.}
\be
\label{eq:2.13ciccia}
\mathcal{F}=-\kappa~\mathrm{vol}(\Omega_{(1)}).
\ee
The coupling constant  $\kappa$  in eq. \eqref{eq:2.13ciccia} counts the unit of magnetic flux through the relation $\kappa =\frac{\pi k}{\sqrt{\lambda}}$.

For a single circular Wilson loop of radius $R$ (parallel to the defect) we expect to find two competing classical string solutions. One is  the usual {\it spherical} dome anchored to the circle  on the boundary of $AdS_5$,
\be
\label{domea}
y^2(\sigma)+r^2(\sigma)=R^2 \ \ \ \  \ \ \ \ \ \phi=\tau,
\ee 
which, however, does not move in the $S^5$. Alternatively, we can consider a second extremal surface
stretching from the boundary to the D5-brane.  This former  is supposed to control the strong coupling behavior of this observable
when  $\frac{L}{R}\gg 1$, while the latter  is expected to dominate the dynamics  in the opposite regime, $\frac{L}{R}\ll 1$.
To determine the second class of extremal surfaces, following \cite{Aguilera-Damia:2016bqv}, we shall postulate  the following ansatz
\be
\label{ansatza}
y=y(\sigma),\quad \quad r=r(\sigma),\quad\quad \phi=\tau,\quad\quad x_3=x_3(\sigma)\quad\mathrm{and}\quad \theta=\theta(\sigma),
\ee
for which the usual Polyakov action in conformal gauge  reduces to 
\be
\label{eq:2.7}
S=\frac{\sqrt{\lambda}}{4\pi}\int d\tau d\sigma \frac{1}{y^2}(y^{\prime 2}+r^{\prime  2}+r^2+x_3^{\prime 2}+
y^2 \theta^{\prime 2})
\ee
The Eulero-Lagrange equation of motion for the action \eqref{eq:2.7} must be paired with the Virasoro constraint 
\be
y^{\prime 2}+r^{\prime  2}+x_3^{\prime 2}+
y^2 \theta^{\prime 2}=r^2.
\ee 
At the boundary of $AdS_5$, which is approached when $\sigma\to 0$, the 
usual Dirichlet boundary conditions must be imposed:
\be
y(0)=0,\quad\quad r(0)=R,\quad\quad x_3(0)=L \quad \mathrm{and}\quad \theta(0)=\chi.
\ee
We have also a second set of boundary conditions to be obeyed where the surface intersects
 the {\it probe}  D5 brane.  We must require  that
\begin{subequations}
\label{eq:2.10}
\begin{align}
\label{eq:2.10a}
C_1\equiv y(\tilde\sigma)-\frac{1}{\kappa} x_3(\tilde\sigma)&=0,\quad\quad\theta(\tilde\sigma)=\frac{\pi}{2},\\
\label{eq:2.10b}
C_2\equiv y^\prime (\tilde\sigma)+{\kappa} x^\prime_3(\tilde\sigma)&=0,\quad\quad C_3\equiv r^\prime(\tilde\sigma)=0,
\end{align}
\end{subequations}
where $\tilde \sigma$ is the maximum value of  $\sigma$.  Eqs. \eqref{eq:2.10a}  and  \eqref{eq:2.10b}  simply state that  extremal surface intersects orthogonally  the boundary brane. 

\noindent
Since the coordinates $x_3$ and $\theta$ are ciclic variables in the action \eqref{eq:2.7}, their equations of motions immediately translate into two conservation laws
\be
\label{eq:2.11}
x_3^\prime(\sigma)=-c y^2(\sigma)\quad\quad\mathrm{and}\quad\quad \theta^\prime(\sigma)=j,
\ee
where $j$ and $c$ are two integration constants to be determined. The equations for $y(\sigma)$ and $r(\sigma)$ are instead
\be
\label{eq:2.12}
yy^{\prime\prime}+r^{\prime 2}+r^2-y^{\prime 2}+c^2 y^4=0\quad\quad y r^{\prime\prime}-2 r^\prime y^\prime -y r=0,
\ee
where we have used eqs. \eqref{eq:2.11} to eliminate the dependence on $x_3$.
The conservation laws eq. \eqref{eq:2.11} also allow us to eliminate the dependence  on $\theta$ and $x_3$  in the Virasoro constraint.
We get
\be
\label{eq:2.13}
{\cal V}(\sigma)\equiv\frac{r^2 -y^{\prime 2}-r^{\prime  2}}{y^2}-c^2 y^2=j^2.
\ee 

\subsection{General solution  for  the connected extremal surface}
\label{subsection:2.1}
First we solve eq. \eqref{eq:2.11} for $\theta$ 
\be
\theta(\sigma)=j\sigma+\chi
\ee
where we used the b.c. $\theta(0)=\chi$. The second boundary conditions $\theta(\tilde\sigma)=\frac{\pi}{2}$ determines the maximum value  $\tilde \sigma$ of the  world-sheet coordinate $\sigma$:
\be
\label{eq:2.15}
\tilde \sigma=\frac{1}{j}\left(\frac{\pi}{2}-\chi\right).
\ee
Next we  focus our attention on the AdS {\it radial} coordinate $y(\sigma)$ and on $r(\sigma)$ which are determined by the system of coupled  eqs. \eqref{eq:2.12}. To solve it we find convenient to introduce the auxiliary function $g(\sigma)\equiv\frac{r(\sigma)}{y(\sigma)}$.  Then, with the help of eqs. \eqref{eq:2.12} and of the Virasoro constraint $\mathcal{V}(\sigma)$, we find
\be
\label{eq:2.16}
\frac{g^{\prime\prime}(\sigma)}{g(\sigma)}=
1-j^2+2 g^2(\sigma),
\ee
a second order differential equation containing only  $g(\sigma)$,  which can be easily  integrated to get the first integral
\be
\label{eq:2.34}
g'(\sigma)^2+(j^2-1)g(\sigma)^2-g(\sigma)^4=-{\varepsilon_0}-j^{2},
\ee
where the arbitrary integration constant has been parameterized as $-{\varepsilon_0}-j^{2}$ for future convenience.  
This equation can be solved explicitly by quadratures through the method of separation of variables; but for the time being,  we will not need the specific form of $g(\sigma)$. 

To determine
$y(\sigma)$, we can use the Virasoro constraint eq. \eqref{eq:2.13} where we have eliminated $r(\sigma)$ in favor of $g(\sigma)$ and performed the change  of variable
\be
y(\sigma)=\frac{1}{\sqrt{1+g^2(\sigma)} z(\sigma)}.
\ee
We find that the unknown function $z(\sigma)$ satisfies the differential equation
\be
\label{eq:2.19cc}\left(g^2(\sigma )+1\right)^2
   z'^2(\sigma
   )-\varepsilon
   _0 z^2(\sigma )+c^2=0.
\ee
Since the first and the last term in the l.h.s. of eq. \eqref{eq:2.19cc} are strictly positive this equation can admit real solutions if and only if 
$
\varepsilon_{0}\ge 0.
$
 We can now easily integrate eq.  \eqref{eq:2.19cc}  by the method of separation of variables and get
\be
\label{eq:2.20aa}
z(\sigma)= \frac{c}{\sqrt{\varepsilon_{0}}}
\cosh [v(\sigma)-\eta].
\ee
where the function $v(\sigma)$ is defined by
 \be
 \label{eq:2.21aa}
 v'(\sigma)=\frac{\sqrt{\varepsilon_{0}}}{1+g^2(\sigma)}, 
 \ee
combined with the boundary condition  $v(0)=0$.  When deriving eq.  \eqref{eq:2.20aa}  we have taken $c>0$ because  $x_3(\sigma)$ must decrease while $\sigma$ grows (see eq.  \eqref{eq:2.11}). The parameter $\eta$ is  an arbitrary  integration constant.  Then the expressions for the original coordinates ($y$ and  $r$)  in terms of  $g(\sigma)$ and $v(\sigma)$ are   given by
\be
\label{eq:3.22}
y(\sigma)= \frac{\sqrt{\varepsilon_{0}}}{c}\frac{1}{\sqrt{1+g^2(\sigma)}} \textrm{sech} [v(\sigma)-\eta]\quad\quad\mathrm{and}\quad\quad
r(\sigma)= \frac{\sqrt{\varepsilon_{0}}}{c}\frac{g(\sigma)}{\sqrt{1+g^2(\sigma)}} \textrm{sech} [v(\sigma)-\eta].
\ee
Finally the coordinate $x_3$ is obtained by integrating eq.  \eqref{eq:2.11}   with respect to $\sigma$. This can be done only in terms of  the  function  $v(\sigma)$ and we obtain
\be
\label{eq:3.23}
x_3(\sigma)=x_0-\frac{\sqrt{\varepsilon_{0}}}{c}\tanh [v(\sigma)-\eta],
\ee
where $x_0$ is another arbitrary integration constant.

Next we can  exploit the boundary conditions  in $\sigma=0$ and $\sigma=\tilde \sigma$ to determine the different integration constants.   Since $g(\sigma)\simeq 1/\sigma$ close to  $\sigma=0$,  the condition $r(0)=R$ becomes
\be
\frac{\sqrt{\varepsilon_{0}}}{c} \textrm{sech} \eta=R\quad\quad\Rightarrow\quad\quad c=\frac{\sqrt{\varepsilon_{0}}}{R} \textrm{sech} \eta.
\ee
Instead  $x_3(0)=L$ translates into 
\be
L=x_0+R \cosh\eta \tanh\eta\quad\quad\Rightarrow\quad\quad  x_0=L-R \sinh\eta.
\ee
A suitable combination of the remaining three boundary conditions $C_i$ given  in eq. \eqref{eq:2.10}  can be used  to determine  $\eta$ in terms of $L$:
\be
0=\kappa c C_1+\frac{1}{y(\tilde\sigma)} C_2+\frac{r(\tilde\sigma)}{y^2(\tilde\sigma)} C_3= c(R\sinh\eta-L)=0
\quad\quad\Rightarrow\quad\quad
\eta=\textrm{arcsinh}\frac{L}{R}.
\ee
Then we are left with two independent boundary conditions in $\sigma=\tilde\sigma$  to impose, for instance $C_1$ and $C_3$, which can be equivalently written as follows
\be
\tanh \left(\eta
   -v\left(\tilde{\sigma }\right)\right)=\kappa \frac{ \text{sech}\left(\eta -v\left(\tilde{\sigma
   }\right)\right)}{\sqrt{g^2\left(\tilde{\sigma }\right)+1}}\quad {\rm and} \quad \tanh \left(\eta -v\left(\tilde{\sigma }\right)\right)=-\frac{
   g'\left(\tilde{\sigma }\right)}{\sqrt{\varepsilon_{0}} g\left(\tilde{\sigma }\right)}
\ee
The latter can be solved  to determine  $L/R$ as function of  the two integration constant $j^2$ and $\epsilon_0$\footnote{Recall that $\tilde \sigma=\frac{1}{j}\left(\frac{\pi}{2}-\chi\right)$ is not an independent variable}:
\be
\label{eq:2.43}
\textrm{arcsinh}\frac{L}{R}=\eta = v(\tilde\sigma)+\tanh ^{-1}\left(-\frac{1}{\sqrt{\varepsilon_{0}}}
\frac{g^\prime(\tilde\sigma )}{g(\tilde\sigma
   )}\right).
\ee
Then the  remaining boundary condition  expresses the geometric flux $\kappa$ in terms of the same variables
\be
\label{eq:2.36}
\begin{split}
\kappa =-\frac{g'\left(\tilde{\sigma
   }\right)}{   \sqrt{j^2+\epsilon _0-g^{2}\left(\tilde{\sigma
   }\right)}}.
   \end{split}
\ee 
The solution of the boundary conditions can be used to  simplify  the form of the parametric representation eq. \eqref{eq:3.22} and  eq. \eqref{eq:3.23} of the extremal surface. We find
\be
\begin{gathered}
y(\sigma)=\frac{ R\cosh\eta}{\sqrt{1+g^2(\sigma)}} \textrm{sech} [v(\sigma)-\eta]\quad\quad
r(\sigma)=R\cosh\eta \frac{g(\sigma)}{\sqrt{1+g^2(\sigma)}} \textrm{sech} [v(\sigma)-\eta]\\
x_3(\sigma)= -R\cosh\eta\tanh [v(\sigma)-\eta].
\end{gathered}
\ee
Finally  we can integrate eq. \eqref{eq:2.34} to construct the explicit form\footnote{We are using Wolfram notation for the elliptic functions, e.g. $\text{sn}(\sigma,m),~\text{cn}(\sigma,m),~\text{dn}(\sigma,m)\dots$.  Next to  $\text{sn},~\text{cn},~\text{dn}$, we can define their inverse $\text{ns}=\frac{1}{\text{sn}},~\text{nc}=\frac{1}{\text{cn}},\text{nd}=\frac{1}{\text{dn}}$ and their ratios $\text{sc}=\frac{\text{sn}}{\text{cn}}$, $\text{cs}=\frac{\text{cn}}{\text{sn}}$, $\dots$} of the function $g(\sigma)$: 
\be
\label{bobo}
g(\sigma)=\sqrt{\frac{j^2-1}{m+1}}\text{ns}\left(\sqrt{\frac{j^2-1}{m+1}}\sigma, m\right),
\ee
where 
\be
\label{parameters}
m\equiv\frac{j^2-1-\sqrt{\left(j^2+1\right)^2+4
   {\varepsilon_0}}}{j^2-1+\sqrt{\left(j^2+1\right)^2+4
   {\varepsilon_0}}}. 
\ee
 Since $\varepsilon_{0}>0$,  the modular parameter $m$ is real and  spans the interval $(-\infty,0]$.  More specifically,  from the definition eq. \eqref{parameters} we  get two ranges  for $m$ according the value of $j^2$: 
 \be
\label{ineq1}
(a):  ~-\infty<m<-1\quad {\rm and}\quad j^2<1 \quad\quad
(b): ~-1\leq m\leq 0\quad{\rm and} \quad j^2\geq 1.
 \ee
For $j^2=1$, we obtain $m=-1$ independently of the value of $\epsilon_0$. In the following  we find more convenient to replace $\epsilon_0$  with $m$  as a free  parameter by solving   eq. \eqref{parameters}. We get 
\be
\label{eps0}
\varepsilon_{0}=-\frac{\left(j^2+m\right) \left(j^2
   m+1\right)}{(m+1)^2}.
\ee
 The positivity of  the integration constant  $\varepsilon_{0}$  combined  with the bounds eq. \eqref{ineq1}  translates into  the following ranges
 for the new couple of free parameters $(m,j^2)$:
   \be
   \label{ineq2}
   \mbox{\sc region (A):}~
   -1\leq m\leq 0~{\rm and }~  j^2\geq-\frac{1}{m}
   \quad {\rm or} \quad  \mbox{\sc region~(B):} ~ m\leq-1~{\rm and }~
   j^2\leq-\frac{1}{m}.
 \ee
\medskip

\subsection{Allowed regions for the parameters \texorpdfstring{${j}, {m}$}{j,m} }
\label{allowedparameters} 
Since eq. \eqref{eq:2.15} explicitly fixes $\tilde \sigma$ in terms of $j$ and $\chi$,  the next step  is to solve the highly non-linear system of equations  \eqref{eq:2.43} and \eqref{eq:2.36} to determine the last two integration constants  $m$ and $j$ as functions of $L/R, \chi$ and $\kappa$.

To begin with, we shall  try to solve eq. \eqref{eq:2.36} for $j$, or equivalently for  the combination
 \be
 \label{xvar}
  x=\sqrt{\frac{j^2-1}{j^2(m+1)}},
 \ee 
 as a function of $m,\kappa$ and $\chi$ .  
Since 
$\kappa =\frac{\pi k}{\sqrt{\lambda}}\geq 0$  and $g(\sigma)\ge 0$ (being the ratio of two positive coordinates), eq. \eqref{eq:2.36} is solved for real values of the parameters if only if $g'(\tilde\sigma)<0$ and
\be
\label{boundA}
g^{2}(\tilde\sigma)\le -m \frac{(j^2-1)^2}{(m+1)^2}
\ee
The bound \eqref{boundA}  ensures that the quantity under the square root in the denominator of eq. \eqref{eq:2.36}  is non-negative.
If we use eq. \eqref{bobo},
the  positivity of $g(\tilde\sigma)$ and the requirement $g'(\tilde\sigma)<0$ can be translated into   the following bounds for $\tilde\sigma=\frac{1}{j}\left(\frac{\pi}{2}-\chi\right)$
\be
\label{eq:2.46a}
0\le \tilde\sigma\le \sqrt{\frac{m+1}{j^2-1}}\mathds{K}(m).
\ee
In terms of the  auxiliary variable $x$ defined in eq. \eqref{xvar}, they read
\be
\label{eq:2.49}
0\leq x \leq\frac{\mathds{K} \left(m\right)}{\left(\frac{\pi }{2}-\chi \right)}.
\ee
The bounds \eqref{ineq2}   for  $j^2$  translate into  $x\ge 1$ independently of the region.  Thus the range of $x$ is given by
\be
\label{eq:2.49a}
1\leq x\leq\frac{\mathds{K} \left(m\right)}{\left(\frac{\pi }{2}-\chi \right)}.
\ee
 \begin{wrapfigure}[15]{l}{73mm}
\centering
\vskip -.4cm
	\includegraphics[width=.33\textwidth]{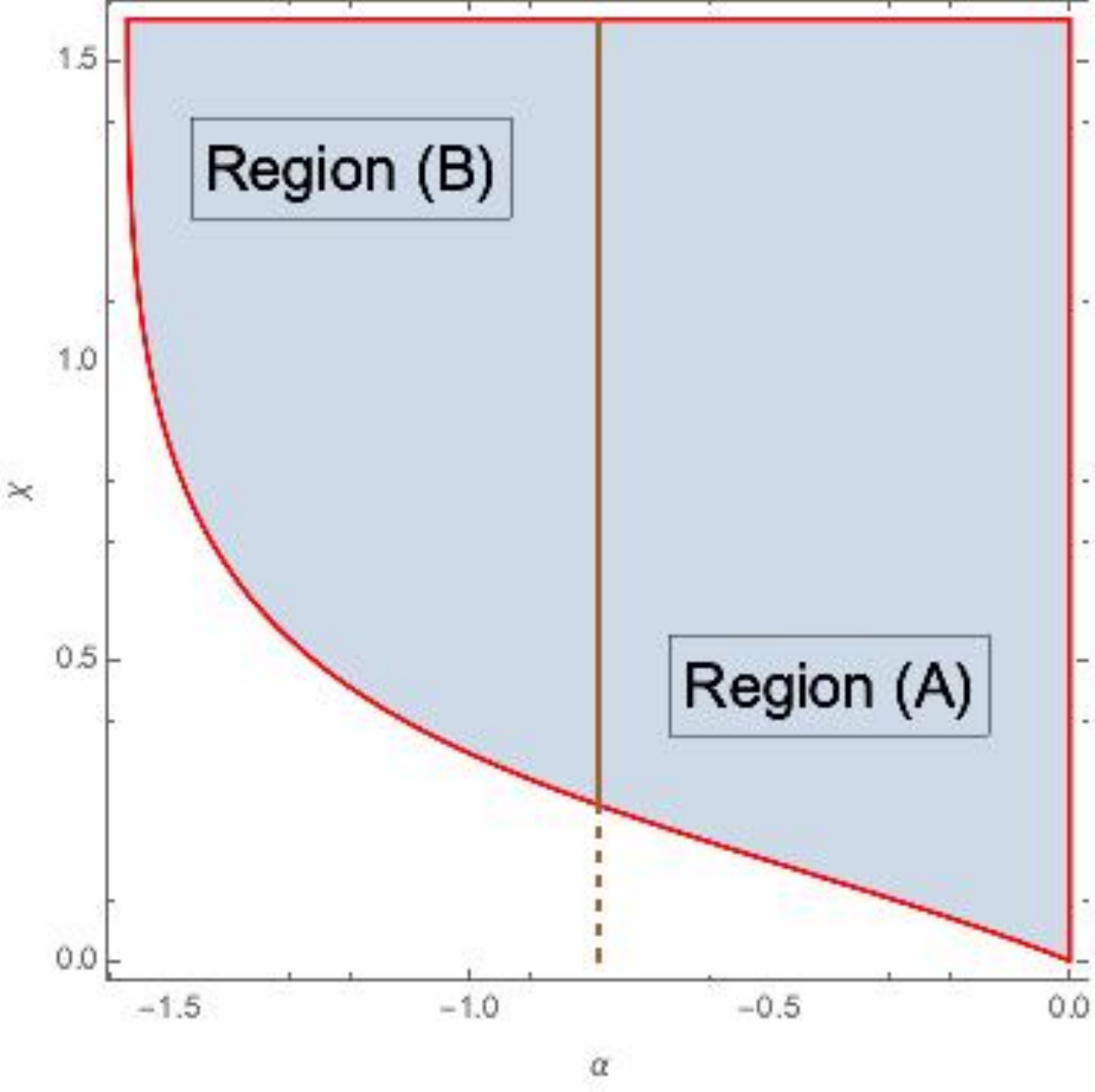}
	\vskip -.3cm
\caption{\label{fig1a} \small The light-blue region  with the red boundary defines the allowed region in the $(\alpha,\chi)$-plane.}
\end{wrapfigure}
In the region (A) the variable $x$    is always less or equal to $\frac{1}{\sqrt{1+m}}$  by construction.  Therefore,
in this region,  we  can refine the bounds  \eqref{eq:2.49a}  as follows
  \be
\label{eq:2.49ab}
1\leq x\leq\text{Min}\left(\frac{1}{\sqrt{1+m}},\frac{\mathds{K} \left(m\right)}{\left(\frac{\pi }{2}-\chi \right)}\right).
\ee
A necessary condition for the existence of solutions of eq. \eqref{eq:2.36} is that
the  intervals  \eqref{eq:2.49a} and \eqref{eq:2.49ab}  are not empty. 
We can solve  this requirement numerically. If we set $m=\tan\alpha$  with $\alpha\in[-\frac\pi2,0]$, 
 the allowed region in the $(\alpha,\chi)$-plane  is the  light blue  area  in fig.\ref{fig1a} bounded by the red line.  The curved boundary  is given by
 \be
\label{chim01}
\chi=\frac{\pi}{2}-\mathds{K}(m)=\frac{\pi}{2}-\mathds{K}(\tan\alpha),
\ee 
namely the pairs $(\alpha,\chi)$ for which the intervals \eqref{eq:2.49a} and \eqref{eq:2.49ab} collapses to a point.

In the above analysis we have neglected  the constraint \eqref{boundA},  which  in terms of $x$ reads
\be
\label{eq:2.50a}
 \text{cn}^2\left(\left.x \left(\frac{\pi }{2}-\chi \right)\right|m\right)\leq- \frac{1}{m}\left(1-\frac{1}{x^2} \right).
\ee 
This inequality  implies  the existence of stronger lower  bound $x_0\ge 1$   for the  unknown $x$.  The value $x_0$ is defined as the value that saturates the inequality \eqref{eq:2.50a},
i.e.
\be
\label{eq:2.50b}
-m \text{cn}^2\left(\left.x_0 \left(\frac{\pi }{2}-\chi \right)\right|m\right)= 1-\frac{1}{x_0^2} ,
\ee
and respects the bounds \eqref{eq:2.49ab} in region (A) and   \eqref{eq:2.49a} in region (B).
 Summarizing, we have the following two ranges for $x$
	  \be
\label{eq:2.51a}
   \mbox{\sc region (A):}  \quad
x_0 \leq x\leq\text{Min}\left(\frac{1}{\sqrt{1+m}},\frac{\mathds{K} \left(m\right)}{\left(\frac{\pi }{2}-\chi \right)}\right)  \quad  \mbox{\sc region~(B):} \quad x_0 \leq x\leq\frac{\mathds{K} \left(m\right)}{\left(\frac{\pi }{2}-\chi \right)}.
\ee
However the new bounds  for $x$ does not alter the allowed  region in the $(\alpha,\chi)-$plane\footnote{Requiring that these new intervals are not empty yields the same constraints on $\chi$ and $\alpha$.}  in fig. \ref{fig1a}.

 \begin{wrapfigure}[13]{l}{65mm}
\centering
\vskip -.5cm
	\includegraphics[width=.35\textwidth]{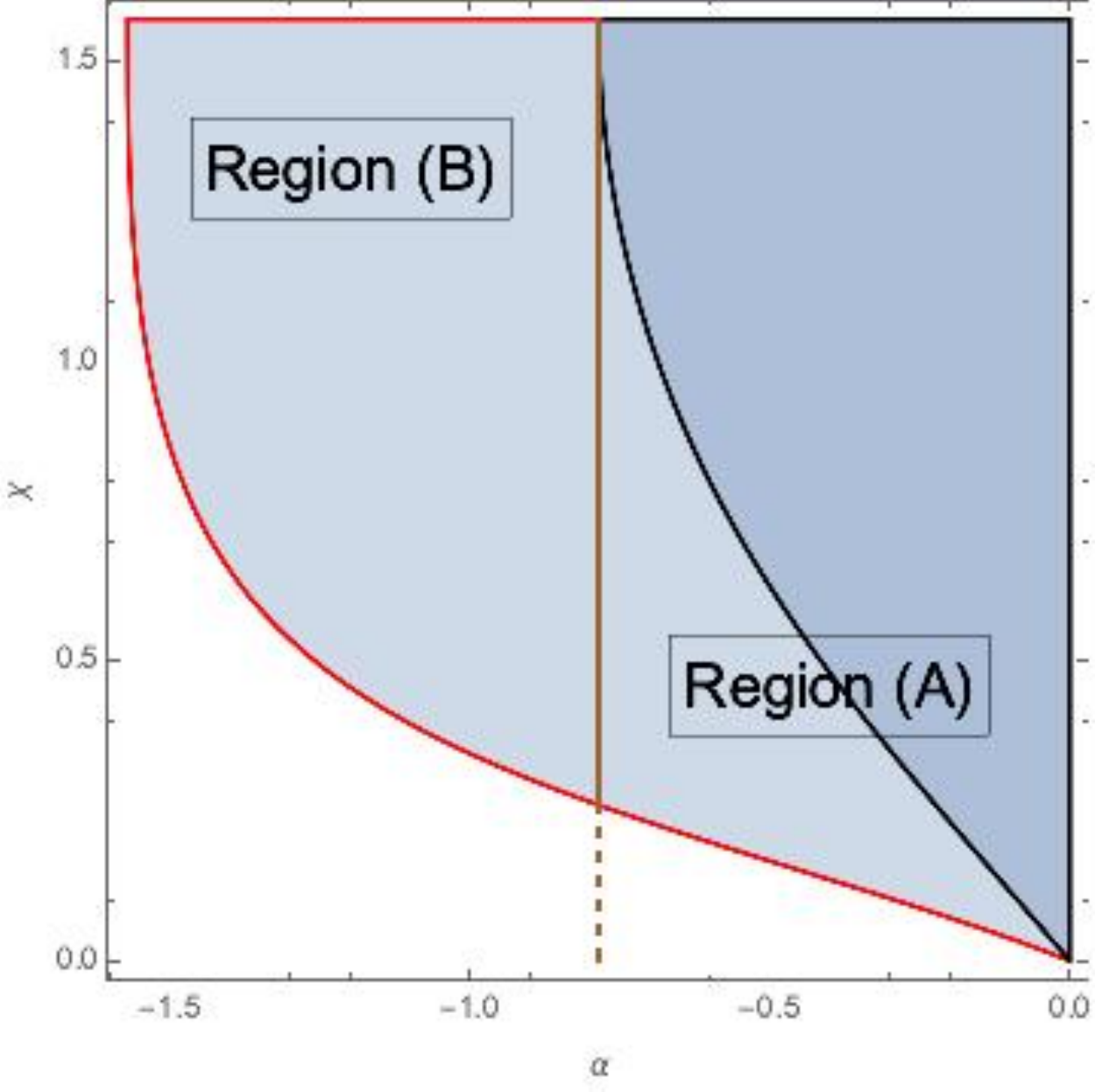}
	\vskip -.3cm
\caption{\label{fig2a} \footnotesize  The inequality \eqref{cico} holds in the light-blue region on the left of the black-line.}
\end{wrapfigure}
Next  we shall analyze how the value of the  flux $\kappa$ may change (in particular reduce) the allowed  region.   Given $m$ and $\chi$, eq.  \eqref{eq:2.36} 
is solved for $\kappa=0$ if we take  $x= \frac{\mathds{K}\left(m\right)}{\frac{\pi }{2}-\chi }$ (namely the value for which $g'(\tilde\sigma)=0$).
 However, in the region (A) this $x$  is an acceptable solution if and only if 
   \be
  \label{cico}
   \frac{\mathds{K}\left(m\right)}{\frac{\pi }{2}-\chi }\leq \frac{1}{\sqrt{1+m}}.
  \ee
 The inequality \eqref{cico} is obeyed in the ligth-blue region on the left of the black curve in fig. \ref{fig2a}. In the darker region on the right of the black curve we cannot solve eq. \eqref{eq:2.36} for arbitrary small value of $\kappa.$
   \noindent
 \begin{wrapfigure}[14]{r}{75mm}
\centering
	\vskip .0cm
	\includegraphics[width=.42
	\textwidth]{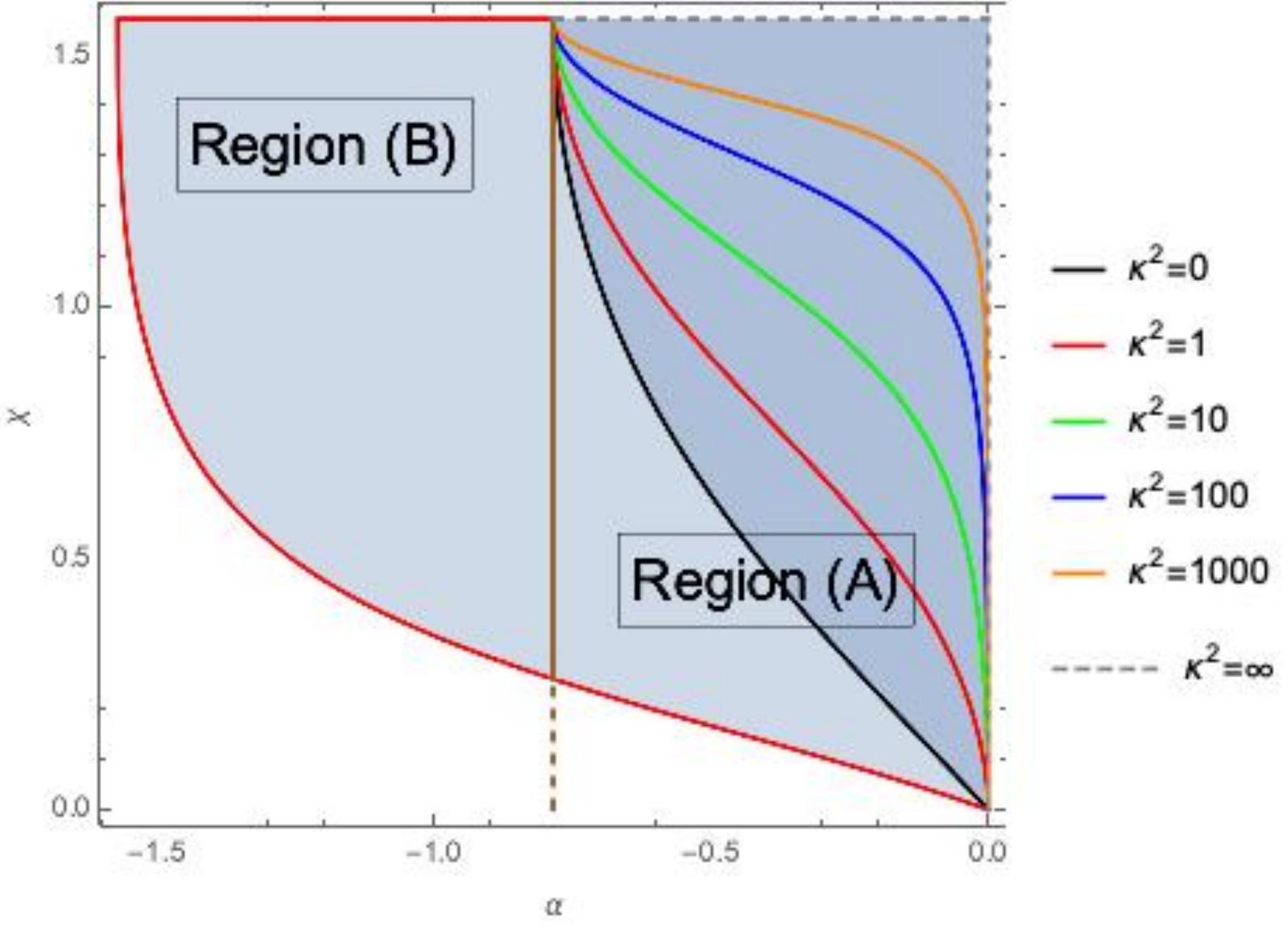}
	\vskip -.3cm
\caption{\label{fig4a} \footnotesize  The coloured curves inside the darker region correspond to different values of $\kappa^2$ . The allowed region for $m$ for fixed value of $\kappa^2$ is the  one  on the left of  the relevant coloured curve. The region becomes larger when we increase the flux.}
\end{wrapfigure}

 We can equivalently  reformulate this obstruction as follows. We fix 
 the flux $\kappa$  and the angle $\chi$ and we increase   $m$ starting from  its lower bound $-1$ in the region (A).  
  We will reach a critical value $m_c$ 
       such that  eq. \eqref{eq:2.36}  is solved by the largest acceptable value for $x$, i.e. $x_c=\frac{1}{\sqrt{1+m_c}}$. Then, for $m\geq m_c$, there is no solution of   eq.  \eqref{eq:2.36} in the interval  \eqref{eq:2.51a}.
      Therefore, given $\kappa$ and $\chi$,  the set of allowed parameters is further restricted by the requirement the l.h.s. of  \eqref{eq:2.36} must be less than $\kappa$ when $x=\frac{1}{\sqrt{1+m}}$:
\be
\label{eq:2.53a}
 \text{sn}\left(\frac{ (\pi -2 \chi )}{2 \sqrt{m+1}}|m\right)\geq\frac{\sqrt{2}}{\sqrt{m+1+\sqrt{(m-1)^2-4 m \kappa ^2}}},
\ee
$m_c$ is the value of $m$ that saturates the inequality \eqref{eq:2.53a}.
In fig. \ref{fig4a}  we have plotted the curve defined by the equality in eq. \eqref{eq:2.53a} for different values of $\kappa$.  Given a specific value of $\kappa$,  the allowed region is the ligth-blue one on the left of the corresponding curve. This region increases when  $\kappa$ grows and we recover the entire region (A)
when $\kappa=\infty.$

\noindent
The  critical value $m_c$ possesses a simple  geometrical interpretation.  In fact we can easily check that the distance $L/R=\sinh\eta$ vanishes as  $m\to m_c$ (see sec. \ref{subsect:3.1}), namely the Wilson loop touches the defect at $m=m_c$ and the solution stops to exist.

The red curve in fig. \ref{fig4a}, which is  the exterior boundary of  the allowed region,  corresponds  to $m=-1/j^2$, i.e. to  $c=0$.  
For this particular choice of the parameters,
our  solutions   coincide with the ones  previously discussed   in \cite{Aguilera-Damia:2016bqv}.  In fact
 our functions simplifies to
\be
\label{eq:2.61a}
\begin{split}
g(\sigma)=&j~\text{ns}\left(j\sigma\left|-\frac{1}{j^2}\right.\right)=\\
=&\sqrt{1+j^2} \text{ds}\left(\left.\sqrt{1+j^2} \sigma\right | \frac{1}{1+j^2}\right)
\end{split}
\ee
and
\be
\label{eq:2.61b}
\begin{split}
h(\sigma)=&\frac{ 1}{R}g(\sigma )\sqrt{1+\frac{1}{g(\sigma )^2}}=\frac{\sqrt{1+j^2}}{R} \text{ns}\left(\sqrt{1+j^2} \sigma\left |\frac{1}{1+j^2}\right.\right),
\end{split}
\ee
where we use the modular properties of the elliptic trigonometric functions.  The solutions  \eqref{eq:2.61a}  and   \eqref{eq:2.61b} are easily seen to be identical to the ones constructed in \cite{Aguilera-Damia:2016bqv}. 

\noindent
Some specific comments are in order for the  two extremal points $\chi=0$  (BPS configuration) and $\chi=\frac{\pi}{2}$.
\paragraph{$\chi=0$ case:} For this choice of the angle governing the coupling of the scalars,  the admissible region for $m$ shrinks to a point, $m=0$ (see fig. \ref{fig4a}). Consequently the integration constant  $j^2$, which must be always greater than  $-1/m$, diverges and  $\tilde \sigma=\frac{\pi}{2j}$ vanishes. In other words, the space of parameters collapses to a point and no regular connected solution exists for the BPS configuration.  
\paragraph{$\chi=\frac{\pi}{2}$ case:}
The case $\chi=\frac{\pi}{2}$  will be discussed in details in sec. \ref{pidiv2}.  At variance with the other values of $\chi$  the disconnected solution cannot exist for all distances: in fact when $\frac{L}{\kappa R}=1$ the disconnected solution touches the brane. If this would happen before the connected solution starts dominating, the phase transition from the disconnected to the connected solution would become of order $0$.

\section{The structure of the solutions}
\label{section:structure}
\subsection{The distance from the defect}
\label{subsect:3.1}
Once we  solved eq. \eqref{eq:2.36} to obtain $x$  (and thus $j$) in terms of $\chi,\kappa$ and $m$,   the distance 
from the defect can be computed through  eq. \eqref{eq:2.43}.  An analytic  expression of this quantity in terms of elliptic integral 
of the third kind is given in appendix \ref{distance}.

The goal of this section is to determine when we can invert eq. \eqref{eq:2.43} and  determine the last integration constant $m$ as a function of the dimensionless distance $L/R$, $\chi$ and $\kappa$.   If we keep fixed the last two quantities, the dependence  of $\frac{L}{R}=\sinh\eta$ on $m$ is monotonic (and thus invertible) if $\frac{\partial\eta}{\partial m}$ does not change sign. 
 We can obtain a compact expression for this derivatives in two steps. First we take the derivative of eq.
 \eqref{eq:3.1}  
 with respect  to $m$: the final result contains a pletora of elliptic trigonometric functions and second elliptic integrals 
  $\mathds{E}(\sqrt{n} \tilde\sigma,m)$.  We can eliminate the last
           dependence by exploiting the derivative of eq. \eqref{eq:2.36} for the flux. The final expression is  relatively simple
    \be
     \label{eq:3.6}
  \left.  \frac{\partial \eta}{\partial m}\right|_{\chi,\kappa}=\frac{\left[\frac{n^2(n+1)  \left(m n+1\right)}{1-m}-m \left(1-m\right) \left(\partial_m n-\frac{n
   (n+1)}{1-m}\right)^2\right] \left(\frac{2 g(\tilde\sigma ) g'(\tilde\sigma )}{2 m n^2 g(\tilde\sigma
   )^2+g(\tilde\sigma )^4-m j^2 n^2}+\frac{\pi -2 \chi }{j^3}\right)}{4 n \sqrt{-(n+1)
   \left(m n+1\right)}},
       \ee  
where the derivative is taken at  constant $\chi$ and $\kappa$  and
\be
n\equiv\frac{j^2-1}{m+1}.
\ee
In eq.  \eqref{eq:3.6} $\partial_m n$ denotes the derivative of $n$ with respect to  $m$. It is not difficult to check that the second factor in 
eq. \eqref{eq:3.6} is  always positive  in the range $n\leq g^2(\sigma)\leq -m n^2$. Thus the sign of eq. \eqref{eq:3.6} is entirely controlled by the  factor  between square brackets.

In subsec. \ref{allowedparameters}, we argued that  we can find a value $m_c$  of $m$ such that  no 
solution exists for $m>m_c$ for fixed $\chi$ and $\kappa$.  This critical value  $m_c$   solves  (see eq. \eqref{eq:2.53a})
\be
\label{mc}
\text{sn}\left(\frac{ (\pi -2 \chi )}{2 \sqrt{m_c+1}}|m_c\right)=\frac{\sqrt{2}}{\sqrt{m_c+1+\sqrt{(m_c-1)^2-4 m_c \kappa ^2}}}.
\ee
We can now easily expand $n$ around $m_c$ and at the leading order we find\footnote{The explicit form of $c_0$ is not relevant at the moment.}
\be
\label{nmc}
n=\frac{x^2}{1-(m+1) x^2}=\frac{c_0}{m_c-m}+O(1)\ \ \ \ \mathrm{with}\ \ \ \  0<c_0<1,
\ee 
namely $n$ blows up at $m=m_c$.
The bounds  on the constant  $c_0$  are equivalent  to the fact that $x$ decreases for $m<m_c$. The   combination $\sqrt{n}\tilde\sigma$ is instead finite  when  $m$ approaches $m_c$ (i.e.  $\sqrt{n}\tilde\sigma\mapsto \frac{\pi-2\chi}{2\sqrt{1+m_c}}$). In this limit it is quite straightforward to show that  $\eta$    vanishes. In fact the argument of $\tanh ^{-1}$ in eq. \eqref{eq:2.43} behaves   as $1/\sqrt{n}$, while
$v(\tilde \sigma)$ vanishes as  $1/\sqrt{n}$ (see appendix \ref{distance}).
 \begin{figure}[t]
                      	\begin{subfigure}{.5\textwidth}
	\includegraphics[width=.95\textwidth]{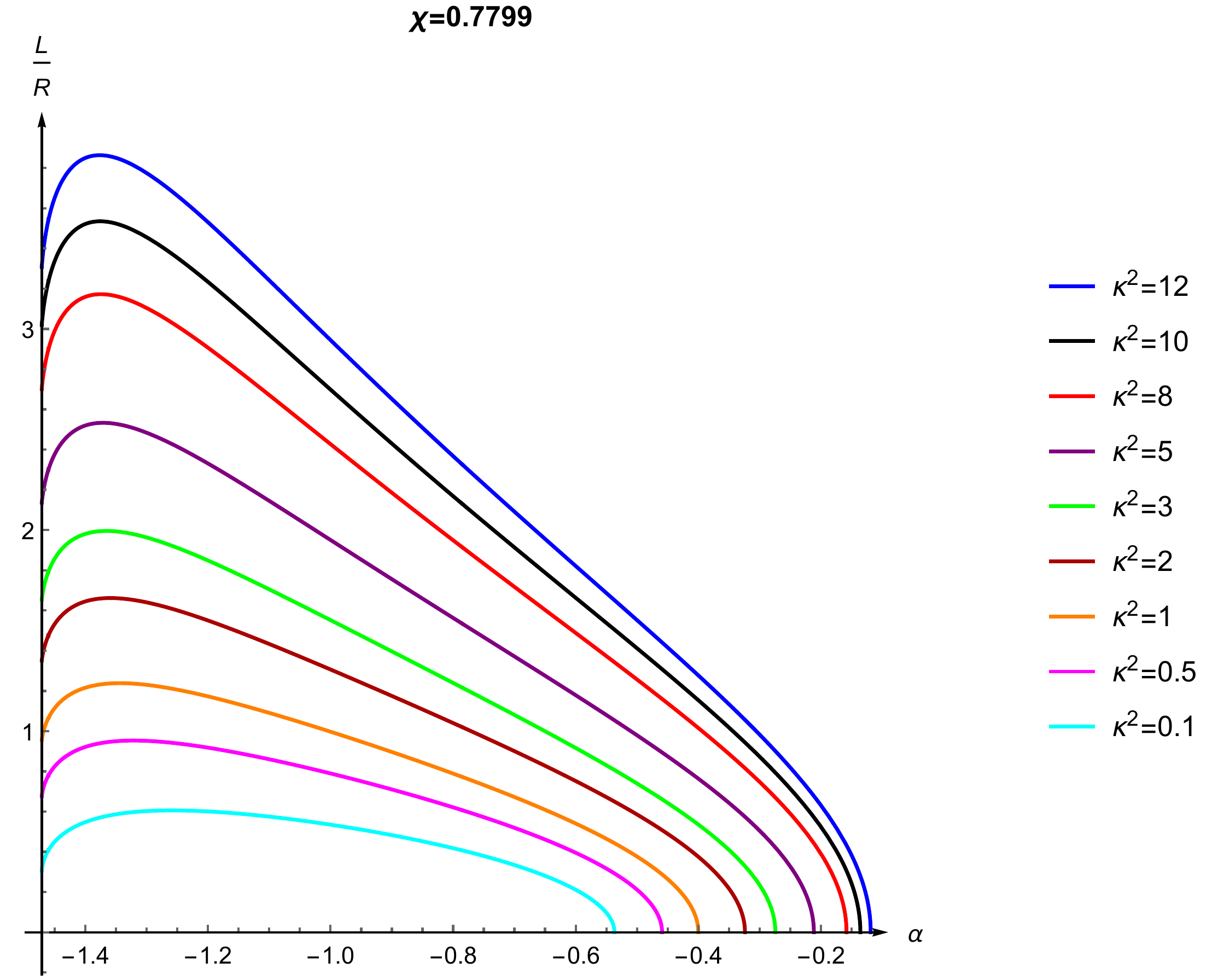}
	\subcaption[\tiny (a)]{}
	\end{subfigure}
                      	\begin{subfigure}{.5\textwidth}
		\includegraphics[width=.99\textwidth]{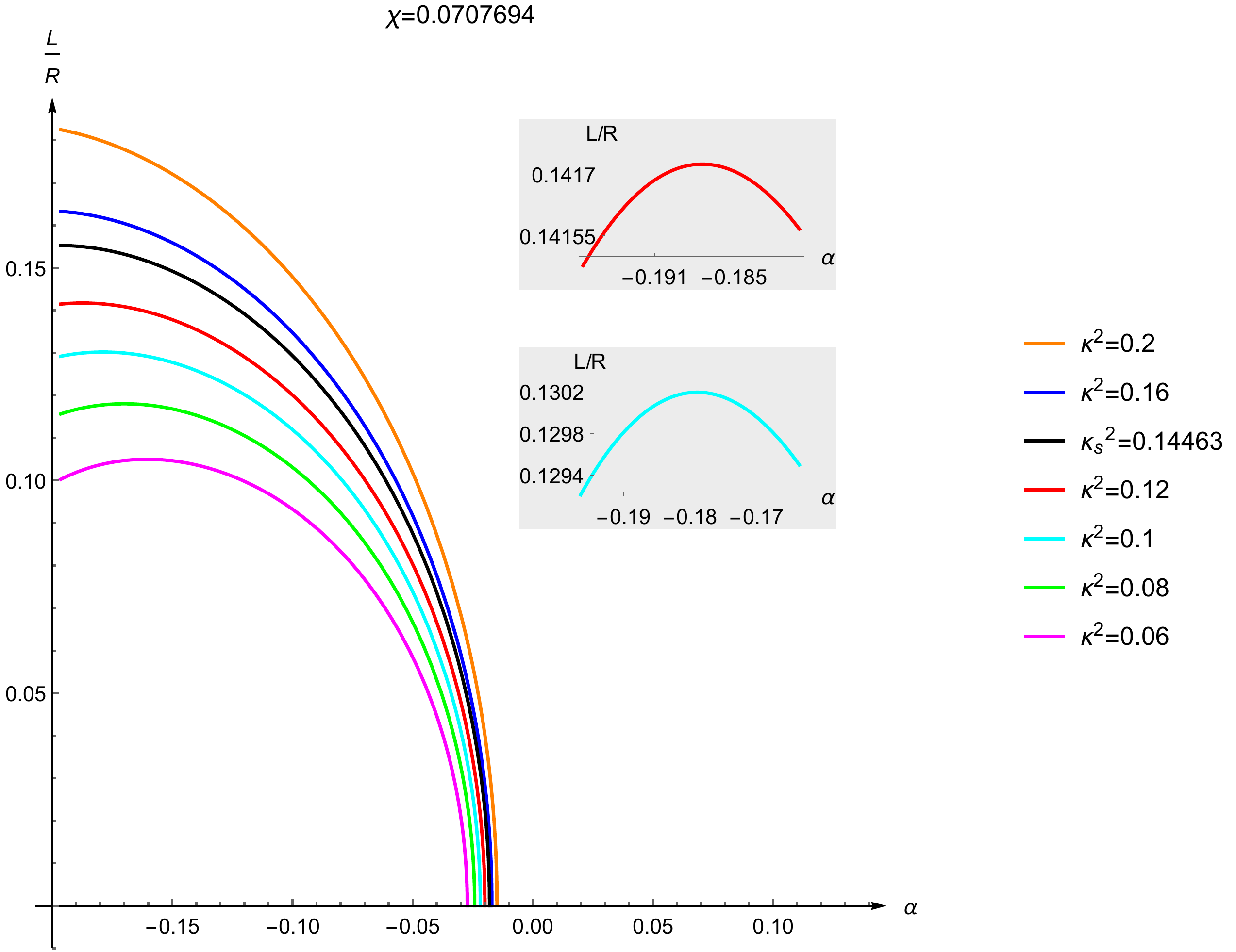}
				\subcaption[\tiny (b)]{}
			\end{subfigure}
		\caption{\label{fig.6} \footnotesize We have plotted the distance $L$ as a function of $\alpha=\arctan(m)$. In (a) we have chosen an
			angle $\chi>\chi_s$: the different curves corresponds to different values of the flux.  All the curves display a maximum. In (b) we 
			have chosen an  $\chi<\chi_s$. When we are above the critical flux $\kappa>\kappa_s$ (the black curve), the curves are monotonic.  On the contrary, as  $\kappa$ becomes smaller than $\kappa_s$ (i.e. we are below the black curve) again a maximum appears.}
\end{figure}

If we use the above behavior of $n$  close to $m_c$, we can also check that the derivative of $\eta$ in $m_c$  and thus of the distance diverges  to
$-\infty$  as  $\sqrt{n}$. This is consistent with  the behavior of the curves  plotted in fig. \ref{fig.6}. When we decrease $m$ the parameter $\eta$ increases, i.e.  we are moving away from the brane. To understand if this behavior takes place for all the range spanned by  $m$
at fixed $\kappa$ and $\chi$, we shall investigate the sign of  $ \frac{\partial \eta}{\partial m}$ when we reach the other boundary of the allowed region, namely the red curve in fig. \ref{fig1a}. The value $m_0$ lying on this second boundary is determined only by the angle $\chi$ and it satisfies
\be
\label{chim0}
\chi=\frac{\pi}{2}-\mathds{K}(m_0).
\ee
Then the derivative of $\eta$ with respect to $m$  computed at $m=m_0$ is given by (see app. \ref{Expansionm0} for a derivation of this result)
\be
\label{derm0}
\begin{split}
\left.\frac{\partial\eta}{\partial m}\right|_{m=m_0}
   =&\frac{\left(1-m_0\right) \left(\mathds{E}\left(m_0\right)+(m_0-1)
   \mathds{K}\left(m_0\right)\right){}^2+m_0\kappa^2\left(1-\left(\mathds{E}\left(m_0\right)+\left(m_0-1\right)
   \mathds{K}\left(m_0\right)\right){}^2\right)}{2 \kappa 
   \left(1-m_0\right) \left(-m_0\right){}^{3/2} \sqrt{1-\left(\kappa ^2+1\right) m_0}}
   \end{split}
\ee
\begin{wrapfigure}[13]{r}{85mm}
 \vskip -0.1cm
\centering 
	\includegraphics[width=.41
	\textwidth]{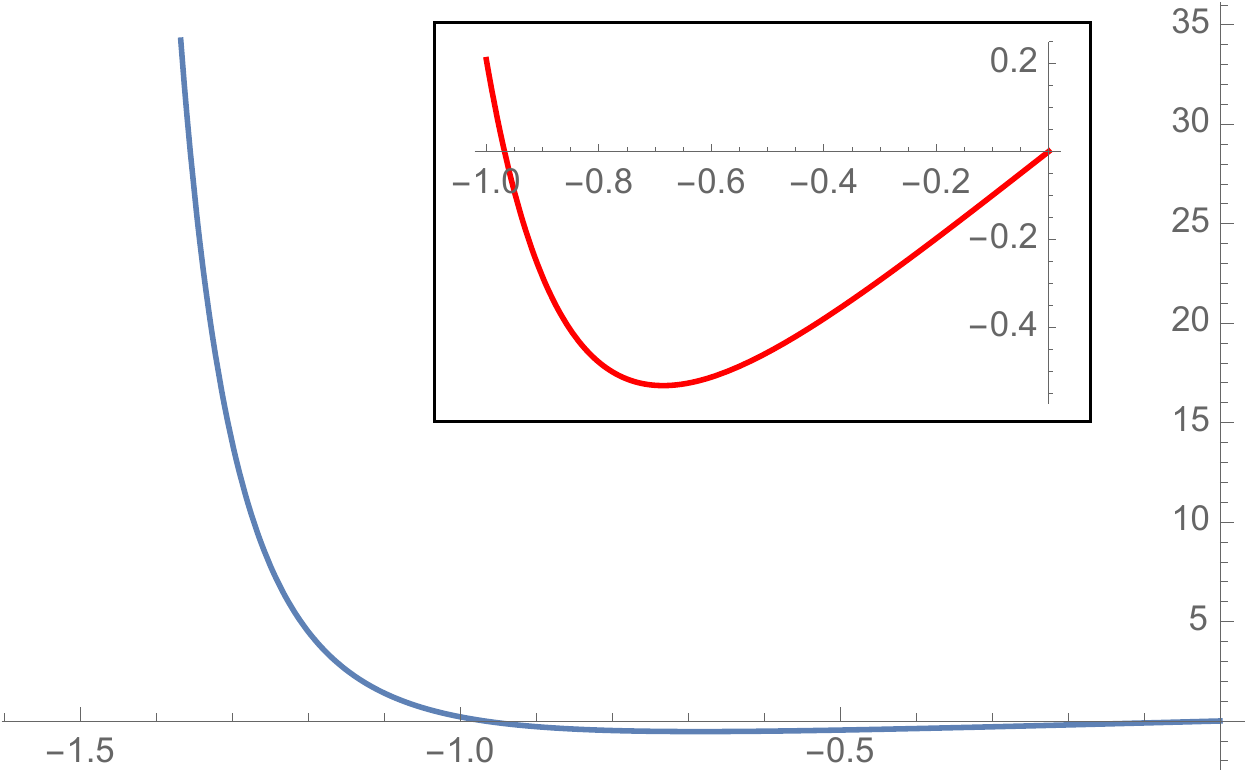}
	\vskip -.3cm
\caption{\label{fig5s} \footnotesize  Plot of the coefficient of $\kappa^2$ in the numerator of eq. \eqref{derm0} 
[we have set $m_0=\tan\alpha\ \textrm{with}\  \alpha \in  (-\pi/2,0]$].  It is negative between zero and and a critical value  $m_s = \tan\alpha_s = -1.45221$. For $m <m_s$ the coefficient is always positive.}
\end{wrapfigure}
The sign of this quantity is controlled by the sign of the coefficient of  the term linear in $\kappa^2$ in the numerator of  eq. \eqref{derm0}
since  the other term is manifestly positive for negative values of $m_0$. In fig.  \ref{fig5s}  we have plotted
the coefficient of  $\kappa^2$.   We recognize that there is  a critical value $m_s\simeq -1.45221$ for which this coefficient vanishes,
namely 
 \be
\left(m_s-1\right)\mathds{K}\left(m_s\right)+\mathds{E}\left(m_s\right)+1=0.
\ee

We can translate $m_s$ into an equivalent critical angle $\chi_s=\frac{\pi}{2} -\mathds{K}(m_s)\simeq 0.331147$.
The  angle $\chi_s$ separates two distinct regions of parameters:
\paragraph{\bf(1):} $\mathbf{m_0 \boldsymbol{\leq} m_s} \ \textrm{(or}\ \boldsymbol{\chi_s\leq\chi\le\frac{\pi}{2}}$).  The coefficient of $\kappa^2$ and thus the derivative  in $m=m_0$ are positive independently of the value of the flux. Since the same derivative diverges to $-\infty$ on the other extremum of the interval $m=m_c$, the dependence of $\eta$ on $m$ cannot be monotonic and we cannot invert  eq. \eqref{eq:2.43}  on the entire range of the allowed $m'$s.   In fact, as illustrated in fig. \ref{fig.6}~(a), when  we decrease $m$ starting from the value $m_c$ defined by  eq. \eqref{mc} the distance from the brane starts increasing, it reaches a maximum and then decreases to the value reached for $m=m_0$.

\begin{wrapfigure}[17]{l}{80mm}
 \vskip .7cm
\centering 
	\includegraphics[width=.41
	\textwidth]{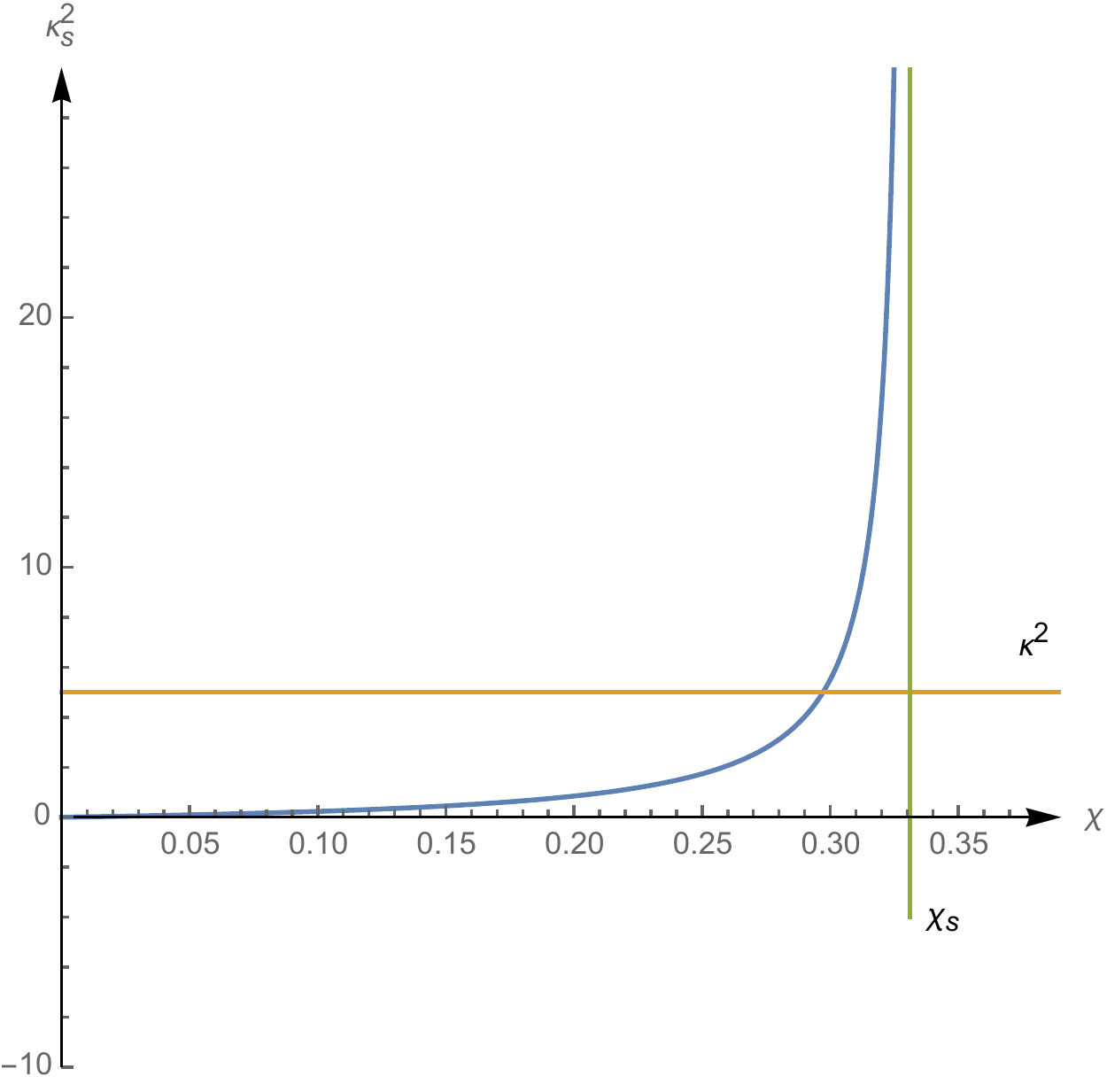}
	\vskip -.3cm
\caption{\label{kappavschi} \footnotesize  The blue curve illustrates the behavior of the critical flux in the interval $[0,\chi_s]$,
where it exists. It is a monotonic function of $\chi$ and it ranges from $0$ to   $\infty$.}
\end{wrapfigure}

\paragraph{\bf(2):} $\mathbf{m_s \boldsymbol{<} m_0< 0} \ \textrm{(or}\ \boldsymbol{0<\chi<\chi_s}).$ 
The derivative of $\eta$ with respect to $m$  computed at $m=m_0$ is always negative unless the flux $\kappa^2$ is below a the critical value $\kappa_s^2$ given by 
\be
\label{criticalflux}
\kappa _s^2= \frac{m_0-1}{m_0 \left(\frac{1}{\left(\left(m_0-1\right)
   K\left(m_0\right)+E\left(m_0\right)\right){}^2}-1\right)}.
\ee
We find more instructive to view this critical flux $\kappa _s^2$ defined in eq. \eqref{criticalflux} as a function of the angle $\chi$ (instead of $m_0$) by exploiting  eq. \eqref{chim0}.  If we  draw the curve $\kappa _s^2(\chi)$ we obtain the blue curve in fig.  \ref{kappavschi}. 
Given the angle $\chi$ (with $0<\chi<\chi_s$) we can determine a critical value of the flux  $\kappa _s$ by means of the plot \ref{kappavschi}. The black curve in fig. \ref{fig.6}~(b)  displays the behavior of the distance with $m$ for the critical value of the flux: it is monotonic and has vanishing derivative at $m=m_0$.  If we choose a flux greater than $\kappa _s$ (the curves above the black one),
the distance  is a monotonic  function of  $m$, namely its derivative never vanishes.  Below this critical value  of the flux (i.e.  the curves below the black one) the derivatives vanishes just once: namely when we decrease $m$ starting from the value $m_c$ the distance  starts increasing, it reaches a maximum and then decreases to the value reached for $m=m_0$.

The presence of a non-monotonic behavior (for a certain range of parameters) is  synonymous of the existence of different branches of solutions. In other words, if either $\chi_s\leq\chi\le\frac{\pi}{2}$ or  $0<\chi<\chi_s$ and $\kappa^2< \kappa _s^2$, we can find value of the distance for which we can construct two different extremal  connected surfaces. We shall come back to this point when we discuss
the area of the extremal surfaces.

In both regions ((1) and (2))  there exists a maximal distance $L_{\rm \tiny max}$ after which the  connected solution stops to exist. When $0<\chi<\chi_s$ and we are above the critical flux $\kappa_s$, $L_{\rm \tiny max}$ is  obtained when we reach the boundary $m=m_0$.  since we are considering the range of parameters for which distance is a monotonic function of $m$. If we use the 
expansion in App. \ref{Expansionm0} and substitute into eq. \eqref{eq:3.1} we find
\be
\label{maxdist}
\eta_{\max}=\tanh^{-1}\sqrt{\frac{\kappa ^2 m_0}{\kappa ^2 m_0+m_0-1}}\ \ \ \ \ \Rightarrow\ \ \ \ \ \ \ L_{\rm\tiny max}=R\sinh\eta_{\tiny\rm max} =R \sqrt{\frac{\kappa ^2 m_0}{m_0-1}}
\ee

\begin{figure}[h!]
 \centering
\includegraphics[width=0.75 \textwidth]{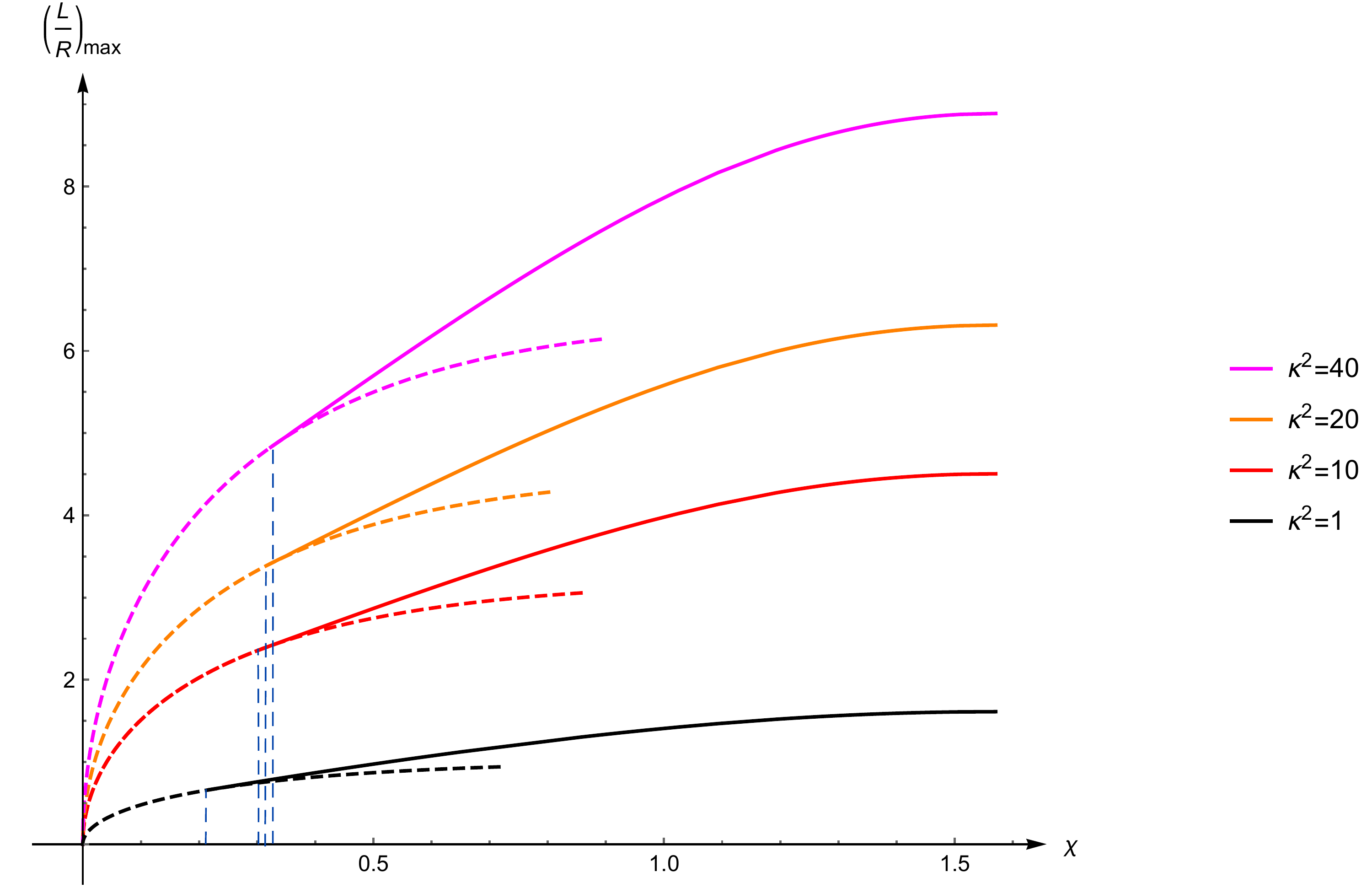}
	\caption{\footnotesize \label{figdistanza} The dashed curves are given by eq. \eqref{maxdist} describe the distance for $\chi<\chi_{\kappa^2 }$ while the continuous ones are valid  for $\chi>\chi_{\kappa^2 }$. Different colors  corresponds to different fluxes. We see the maximal distance grows both with $\chi$ and $\kappa^2$. We have drawn the dashed curves also for values greater than  $\chi_{\kappa^2}$ to
 show that the merging of the two branches is continuous with its first derivative.}
	\end{figure}
If we are below the critical flux or the angle $\chi$ is in the range $(\chi_s,\frac{\pi}{2})$ the maximal distance from the brane cannot  be determined analytically, but only numerically. In fig. \ref{figdistanza}  we have
plotted this quantity as a function of $\chi$ for different values of the flux $\kappa^2$.  Given 
$\kappa^2$,  we can always find an angle $\chi_{\kappa^2}\in [0,\chi_s]$ for which 
$\kappa^2$ is exactly  the critical flux (see  fig.  \ref{kappavschi}).  For angles in the interval $[0,\chi_{\kappa^2})$,  the maximal distance is given by eq. \eqref{maxdist} and is represented by the dashed curves in fig. \ref{figdistanza}.
 For angles greater than $\chi_{\kappa^2}$,
the behavior of the distance is described by the continuous lines in fig. \ref{figdistanza}.
The merge of the two branches of the distance (dashed and continuous) is continuous with its first derivative as one can show by a direct computation of the left and right derivative in $\chi_{\kappa^2}$\footnote{The derivative of $\eta$ with respect to $\chi$ admits a very simple form in terms of $n$ and $m$
	 \[
\left. \frac{\partial \eta}{\partial \chi}\right|_{\kappa, m}=\frac{(m-1) m \partial_m n+(n+1)(2 m n+1)}{j \sqrt{-(n+1)
   (m n+1)}}.
\]}.
In fig. \ref{figdistanza}, we have drawn the dashed curves also for values greater than  $\chi_{\kappa^2}$ to
 illustrate that two branches are not given by the same function.

The general behavior of the distance displayed in fig.  \ref{figdistanza} is easily summarized: the maximal distance increases with $\chi$ at fixed $\kappa^2$ and increases with $\kappa^2$ at fixed angle, The latter  behavior is expected: in fact, when $\kappa^2$ grows the slope of the 
brane becomes smaller and smaller and the brane is {\it closer}  to the boundary. Therefore the cost in {\it energy} ({\it in area}) is low for a larger interval of the distances.

\subsection{The Area}
\label{subsect:3.2}
The regularized area of the connected minimal surfaces is obtained by  evaluating the Polyakov action on the classical solution $(r(\sigma), y(\sigma))$: 
\be
      \label{eq:3.7}
\mathcal{S}=\frac{\sqrt{\lambda}}{4\pi}\int d\tau d\sigma \frac{1}{y^2}(y^{\prime 2}+r^{\prime  2}+r^2+c^{2} y^4+
j^2 y^2 ).
\ee
We can eliminate the explicit dependence on the integration constants $j$ and $c$ by means of the Virasoro constraint \eqref{eq:2.13}. We find
\be
      \label{eq:3.8}
S=
\frac{\sqrt{\lambda}}{4\pi}\int_{\Sigma} d\tau d\sigma \frac{1}{y^2}(y^{\prime 2}+r^{\prime  2}+r^2+r^2 -y^{\prime 2}-r^{\prime  2} )=
{\sqrt{\lambda}}\int_{{\sigma_\epsilon} }^{\tilde\sigma}d\sigma \frac{r^2(\sigma)}{y^2(\sigma)}=
{\sqrt{\lambda}}\int_{{\sigma_\epsilon} }^{\tilde\sigma} d\sigma  g^2(\sigma),
\ee 
where we have used that the integrand does not depend on $\tau$ to perform the integration over this world-sheet coordinate. 
 Remarkably the area depends only on the function $g(\sigma)$. The integration over $\sigma$ runs
from $\sigma_\epsilon$,  the value of $\sigma$ for which  the minimal surface intersects the plane $y=\epsilon$,  to $\tilde \sigma$, the value of $\sigma$ for which  the minimal surface intersects the boundary brane. 
The lower extremum $\sigma_\epsilon$ is determined by solving $y(\sigma_\epsilon)=\epsilon$ for small $\epsilon$. At the lowest orders in $\epsilon$  we find the following expansion:   \[\sigma_\epsilon=\frac{\epsilon}{R}+\frac{1}{6} (j^2+2) \frac{\epsilon^3}{R^3}+O(\epsilon^4).\]
Next we can easily perform the integration  over the coordinate $\sigma$ in terms of elliptic integral of the second kind and we  get 
\be
      \label{eq:3.9}
\mathcal{S}_{\epsilon}=\left.\sqrt{\lambda n} \left( 
   \sqrt{n}\sigma-\mathds{E}\left(\text{am}\left(\sqrt{n} \sigma
   |m\right)|m\right)-\frac{\text{cn}\left(\sqrt{n} \sigma
   |m\right) \text{dn}\left(\sqrt{n} \sigma
   |m\right)}{\text{sn}\left(\sqrt{n}  \sigma |m\right)}\right)\right|_{\sigma_\epsilon}^{\tilde\sigma}.
\ee
The expected ultraviolet linear divergence arises when we evaluate the primitive in  the lower extremum. Since the term $\epsilon^2$ is absent in the expansion of  $\sigma_\epsilon$, the lower extremum does not give any non-vanishing
contribution  in the limit $\epsilon\to 0$.  The only contribution to the renormalized  area originates from
the upper extremum:
\be
      \label{eq:3.10}
\mathcal{S}_{\rm ren.}=\sqrt{\lambda n} \left( 
   \sqrt{n}\tilde \sigma-\mathds{E}\left(\text{am}\left(\sqrt{n}\tilde \sigma
   |m\right)|m\right)-\frac{\text{cn}\left(\sqrt{n}\tilde \sigma
   |m\right) \text{dn}\left(\sqrt{n} \tilde \sigma
   |m\right)}{\text{sn}\left(\sqrt{n}\tilde  \sigma |m\right)}\right)\equiv\sqrt{\lambda} \mathcal{\hat S}_{\rm ren.}.
\ee 
The behavior of the renormalized action as a function of the distance from the defect  can be investigated by computing its derivative with respect to $m$ at fixed $\kappa$ and $\chi$.  We proceed by following the same two steps performed for the evaluation of $\left.\frac{\partial \eta}{\partial m}\right|_{\kappa, \chi}$ and we get the remarkable identity
\be
\label{eq:3.14s}
\left.  \frac{\partial\hat{\mathcal{S}}_{\rm ren.}}{\partial m}\right|_{\kappa, \chi}=\sqrt{-(n+1) (m n+1)}  ~\left.\frac{\partial \eta}{\partial m}\right|_{\kappa, \chi}
  \ee
namely the derivative of the action and of $\eta-$parameter (the distance) are proportional through a positive definite factor.
Thus the derivative of the $\hat{\mathcal{S}}_{\rm ren.}$ with respect to $\eta$ or equivalently to  $L$ has a very simple form\footnote{This relation immediately implies that ${\mathcal{S}}_{\rm ren.}$ does not depend on $L$ when $c=0$, which apparently contradicts eqs.  (24) and (25) in \cite{Aguilera-Damia:2016bqv}. But this is not case. In fact if we keep $\chi$  and $\kappa$ fixed the area  \cite{Aguilera-Damia:2016bqv} does not change with $L$: it is a function only of $\chi$.}
\be
\left. \frac{\partial\hat{\mathcal{S}}_{\rm ren.}}{\partial L}\right|_{\kappa, \chi}=\frac{1}{R\sqrt{1+\frac{L^2}{R^2}}}\left.\frac{\partial\hat{\mathcal{S}}_{\rm ren.}}{\partial \eta}\right|_{\kappa, \chi}=\frac{\sqrt{-(n+1) (m n+1)} }{R\sqrt{1+\frac{L^2}{R^2}}}=c(L,\kappa,\chi)
\ee
where $c$ is the integration constant appearing in eq. \eqref{eq:2.11} for $x_3$. A similar relation was found in \cite{Kim:2001td} for the potential quark-antiquark at finite temperature. There it was speculated that this kind of relation might enjoy some sort of universality.

Because of eq. \eqref{eq:3.14s} the area and the distance possess  the same behavior
as functions of $m$:  
\begin{itemize}
\item[$\boldsymbol{(S_1)}$] $\boldsymbol{\chi_s\le \chi\leq \frac{\pi}{2}}$  or $\boldsymbol{0< \chi< \chi_s}$ and $\boldsymbol{\kappa\le \kappa_s}$: when we move away from the D$5$-brane by decreasing $m$ from its critical value $m_c$ (corresponding to vanishing distance) both the area and  the distance increase  and reach their maximum value for the same value of $m$. Then  both decrease up to  $m=m_0$ (i.e. the curved red boundary in fig. \ref{fig2a}). 
\item[$\boldsymbol{(S_2)}$]   $\boldsymbol{0<\chi< \chi_s}$ and $\boldsymbol{\kappa> \kappa_s}$: both the area and  the distance monotonically increase when $m$ is lowered from $m_c$ to $m_0$.
\end{itemize}
\begin{figure}[t]
	\centering{\includegraphics[width=15cm]{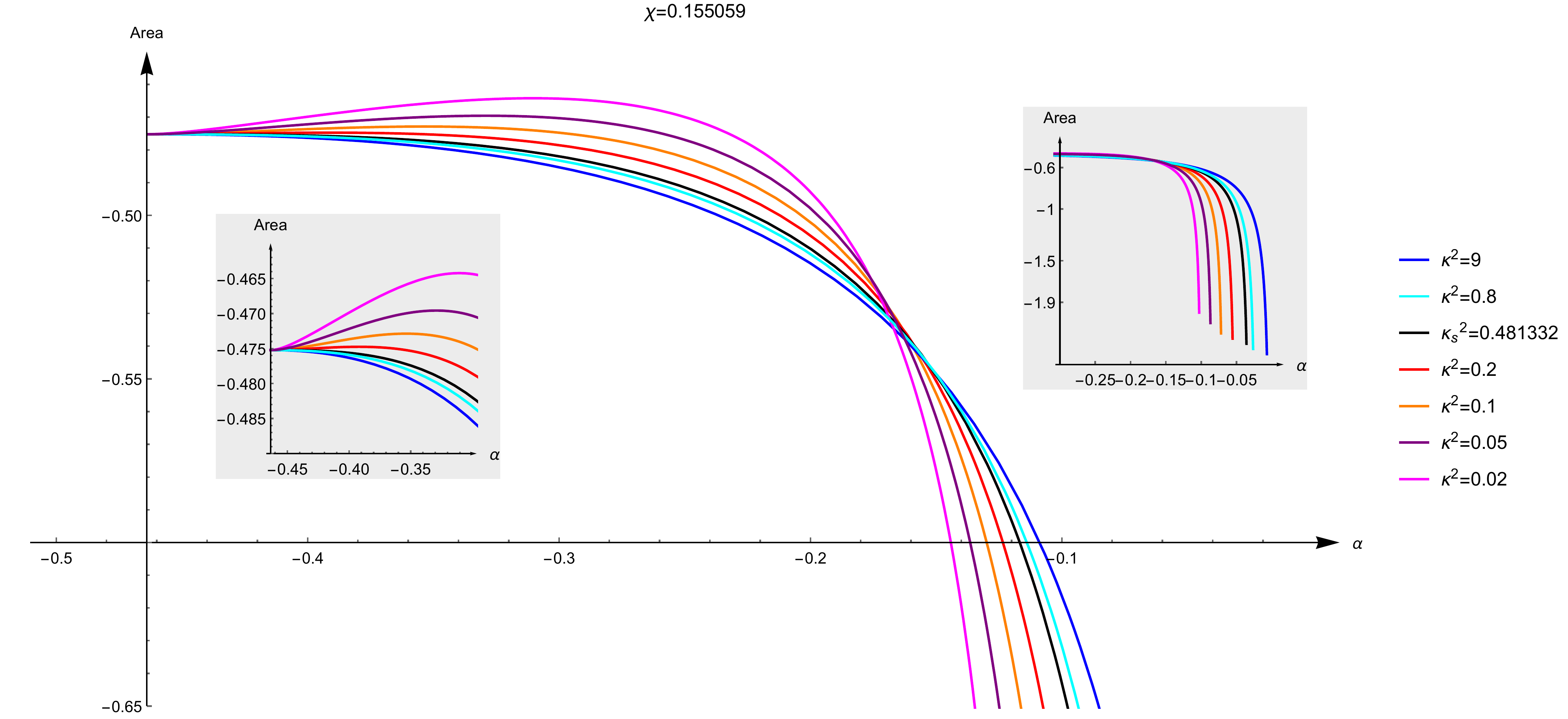}}
	\caption{\footnotesize \label{fig6bis} We have plotted the behaviour  of the area as a function of $\alpha=\arctan (m)$ for a fixed value of the angle $\chi$ (below $\chi_s$) but for different values of the flux. On the left we have zoomed the region close to the red curved boundary in fig. \ref{fig1a}. Independently of $\kappa^2$ all the curves terminate on the same point ($m_0,\chi_0=\frac{\pi}{2}-\mathds{K}(m_0)$). For values of $\kappa^2\le\kappa^2_s$, this curve displays a maximum, instead they are
	monotonic if we are above the critical flux $\kappa_s$. On the  right we have plotted the same curves in the region close to $m_c=\tan\alpha_c$, namely the value for which the distance from the brane vanishes. The area diverges for all value of the flux.
}
\end{figure}
In  fig. \ref{fig6bis} we have plotted  the area as a function of $\alpha\equiv \arctan(m)$ for a fixed angle $\chi=0.155059$ in the range  $0< \chi< \chi_s$.  On the left gray box we have zoomed  on the behavior of the area for different value of the flux in the proximity of  the 
value $m_0$  on the curved red boundary in fig. \ref{fig1a}. Below the critical value of the flux (black curve) determined by eq.\eqref{criticalflux}
all the lines possess a maximum, while above $\kappa^2_s$  they are monotonically decreasing. However, independently of the flux,  all of them terminate in the same point since the boundary value on the red curve in fig. \ref{fig1a} does not depend on the flux.  On the right gray box we have zoomed in the region close to  the brane,  the action always diverges  when $m$ approaches the values $m_c$, obtained by solving eq. \eqref{mc}.

\begin{figure}[t]
	\centering{\includegraphics[width=13cm]{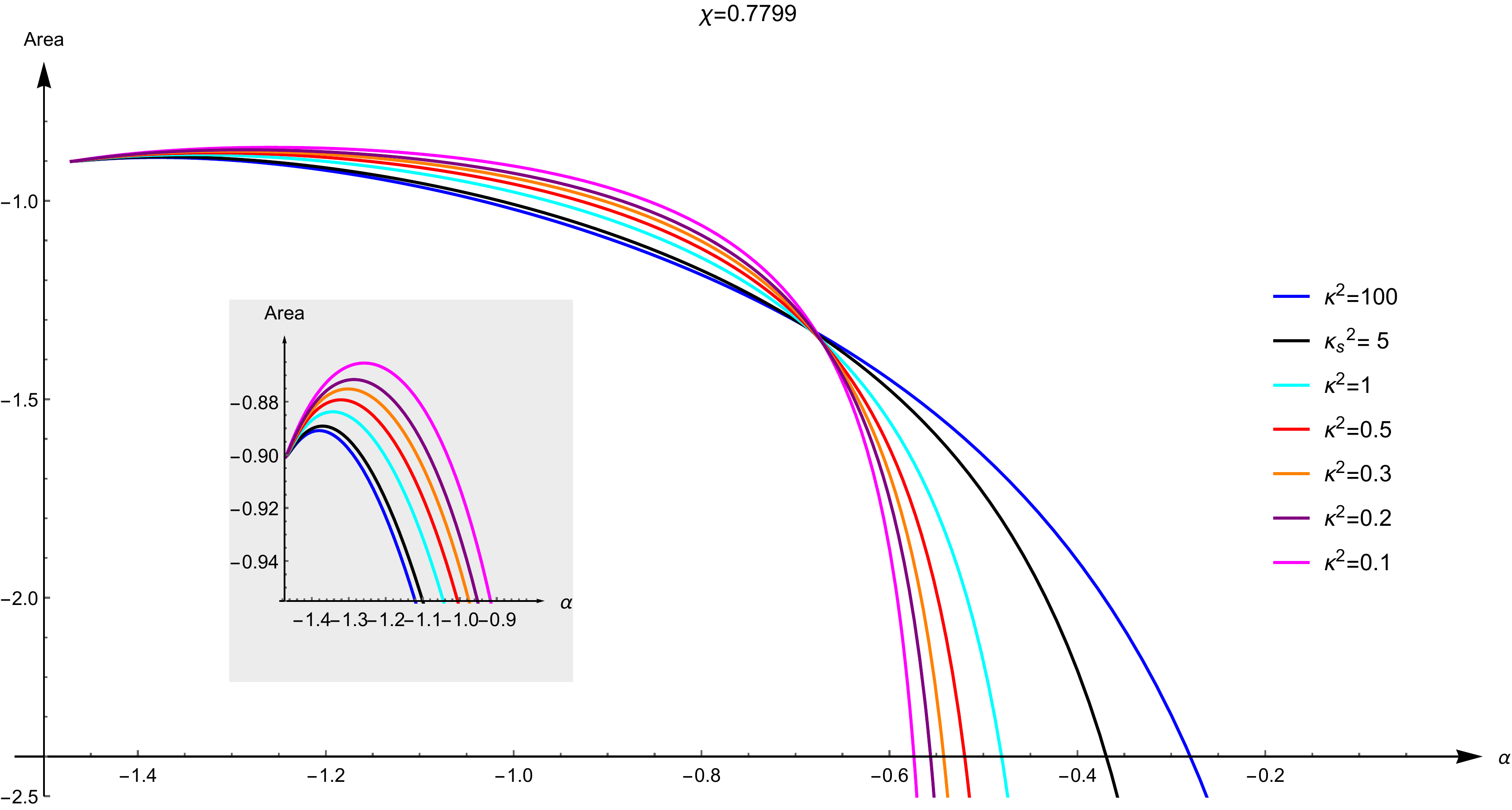}}
	\caption{\footnotesize \label{fig7bis} We have plotted the behaviour  of the area as a function of $\alpha=\arctan (m)$ for a  value of the angle above $\chi_s$  and for different values of the flux. Independently of $\kappa^2$ all the curves terminate on the same point ($m_0,\chi_0=\frac{\pi}{2}-\mathds{K}(m_0)$) and display a maximum. On the  left we have zoomed in the region close to $m_0$ to show the presence of a maximum value for the area.}
\end{figure}

The typical  behavior of the area in the region $\chi_s\le\chi\le\frac{\pi}{2}$ as function of $m$ is displayed in fig. \ref{fig7bis}.  Independently of $\kappa^2$ all the curves terminate on the same point ($m_0,\chi_0=\frac{\pi}{2}-\mathds{K}(m_0)$) and possess a maximum for the same value of $m$ for which the distance does (see left gray box). 

We conclude this section with an amusing observation about the derivatives of the area with respect the two other parameters. It is not
difficult to check that both of them can be rewritten in terms  of the same derivative of $\eta$:
\be
(A_1):\ \ 
  \frac{\partial \mathcal{A}}{\partial \chi}=\sqrt{-(n+1) (m n+1)} ~ \frac{\partial \eta}{\partial \chi}-j
\qquad  (A_2):\ \ 
 \kappa \frac{\partial \mathcal{A}}{\partial \kappa}=\sqrt{-(n+1) (m n+1)} ~ \kappa\frac{\partial \eta}{\partial \kappa}+\frac{g'(\tilde\sigma)}{ g(\tilde\sigma)}
\ee
\begin{figure}[t]
	\centering{\includegraphics[width=15cm]{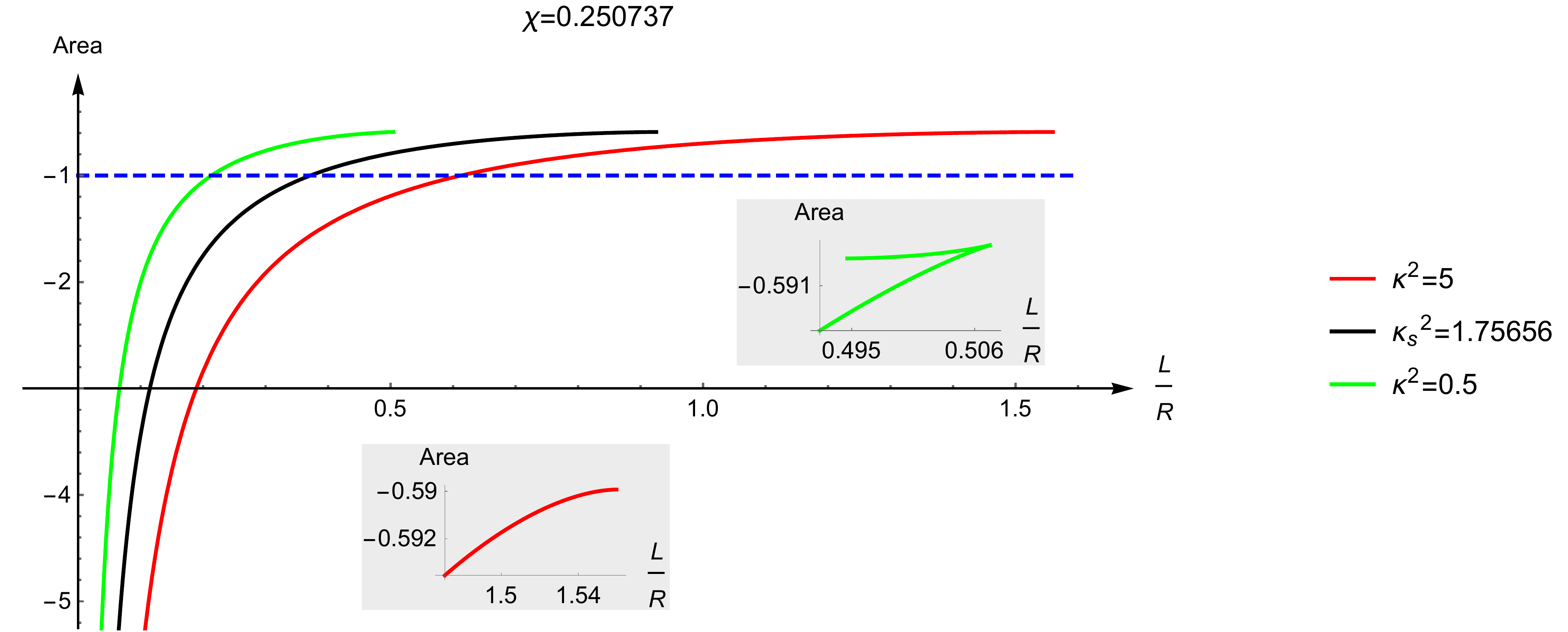}}
	\caption{\footnotesize \label{fig10} This graphic displays the behavior of the area as a function of the  distance $L/R$ for a fixed value of the angle $\chi=0.250737<\chi_s$. In this range there is a critical flux $\kappa_s^2=1.75656$ and the area for this value of the flux is the black curve.  Above the critical flux, the typical behavior is given by the red curve, namely the area is a monotonic function of the distance and the solution stops existing after a maximal value of the distance. Below the critical flux,  the typical behavior is instead described by the green curve. There are two branches of solutions. However the upper one is always subdominant. In this plot we normalized the area functional  so that the area of the dome is $-1$.}
\end{figure}
\subsection{Transition: connected solution vs dome}
\label{subsect:3.3}
To understand when the connected solution becomes  dominant with respect to the spherical dome we have to plot  the area as a function of the distance from the brane. This can be done by exploiting the result of the previous two sections.  We have to distinguish two cases depending on the angle $\chi$ governing the coupling with the scalars:
\paragraph{$\boldsymbol{0<\chi<\chi_s}$:}
Above the critical flux $\kappa_s$ (represented by the black curve in fig. \ref{fig10}) the  area is a monotonic function of the 
distance (see e.g. the red curve in fig. \ref{fig10}). If we are approaching from infinity the connected solution start to exist
at certain maximal distance, which depends on $\chi$ and is given by eq. \eqref{maxdist}. The area is larger than the one of the dome, which, therefore, still dominates.
While we are getting closer the area keeps decreasing and we reach a critical distance where the connected solution and the disconnected one have the same area.  In fig. \ref{fig10} the critical distance is realized when the red curve crosses the blue line. After
this value of the distance, the connected solution becomes  the dominant one: in fact the area keeps decreasing and diverges to 
$-\infty$ when we reach the brane. 

\begin{figure}[t]
	\centering{\includegraphics[width=13cm]{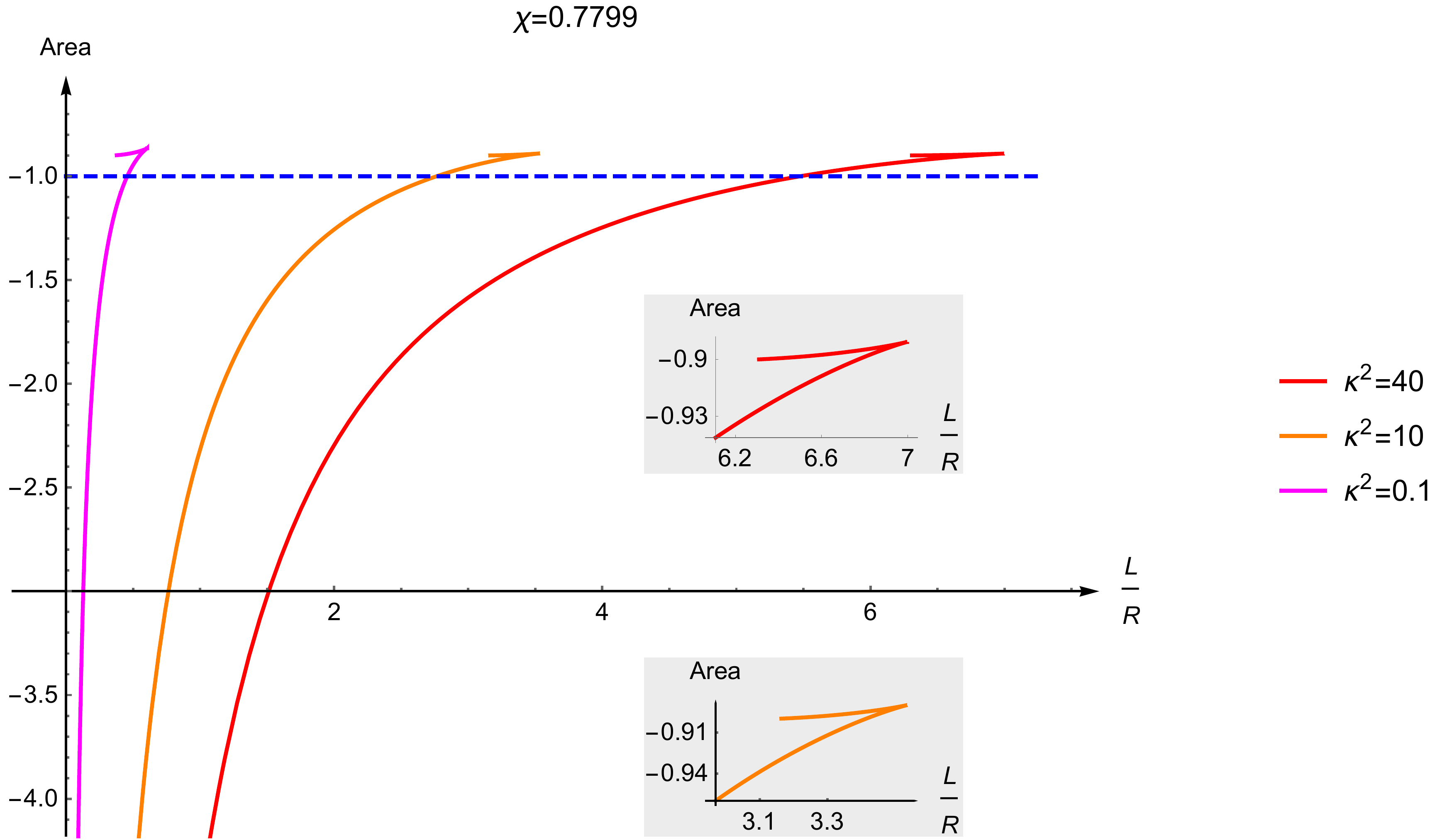}}
	\caption{\footnotesize \label{fig11} This plot displays the behavior of the area as a function of the distance $L/R$ for a fixed value of the angle $\chi=0.7799>\chi_s$. In this range there is no critical flux.  The typical behavior is described by the three  curves (red, orange and purple). There are always two branches of solutions. However the upper one is always subdominant. In this plot we normalized the area functional  so that the area of the dome is $-1$.}
\end{figure}

The typical behavior of the area  below the critical flux $\kappa_s$ is described, instead, by the green curve in fig. \ref{fig10}.  There is still a maximal distance, at which the connected solution starts to exist, but when the distance decreases there are two different branches of solutions: namely we have two connected extremal surfaces with the same angle $\chi$, flux $\kappa$ and distance $L$.  In 
fig. \ref{fig10}, this  is clearly displayed  in the zoom of the region close to the maximal distance.  The upper branch corresponds to the values of the modular parameter $m$ ranging from $m_0$, lying 
on the red curved boundary in fig. \ref{fig1a}, to the value corresponding to the maximal distance.  The lower one is instead
obtained when $m$ runs from  the value corresponding to the maximal distance to $m_c$,  for which the distance from the brane vanishes. The area of the solution in the upper branch, when  it exists, is always  subdominant with respect to the one in the lower branch. Therefore, we can focus on the latter.
\paragraph{$\boldsymbol{\chi_s\le \chi\le \frac{\pi}{2}}$:} The behavior of the area as function of the distance  in this range of angles is displayed in  fig. \ref{fig11}.  The situation is analogous 
to what occurs in the other region below the critical flux. Starting from the maximal distance for which the  connected solution exists, we have two families of extremal surfaces when the distance decreases. However, as we can see in in  fig. \ref{fig11}, the shorter one is always subdominant. The one relevant for us will be the lower branch running from the maximal distance to the brane.
The length of the subdominant branch increases with the flux.

In both regions there is always a value of the distance for which the 
area of the dome is equal to the area of the connected solution and below which the connected solution becomes dominant.  In other words we have a phase transition. In fact  above this critical distance, the dominant solution is the spherical dome and the area is constant (the dashed blue line in fig. \ref{fig10} and \ref{fig11}).  
The transition is of the first order  since the area is continuous but not its first derivative. The critical distance increases with the flux,
as one expects.

\begin{figure}[t]
	\centering{\includegraphics[width=13cm]{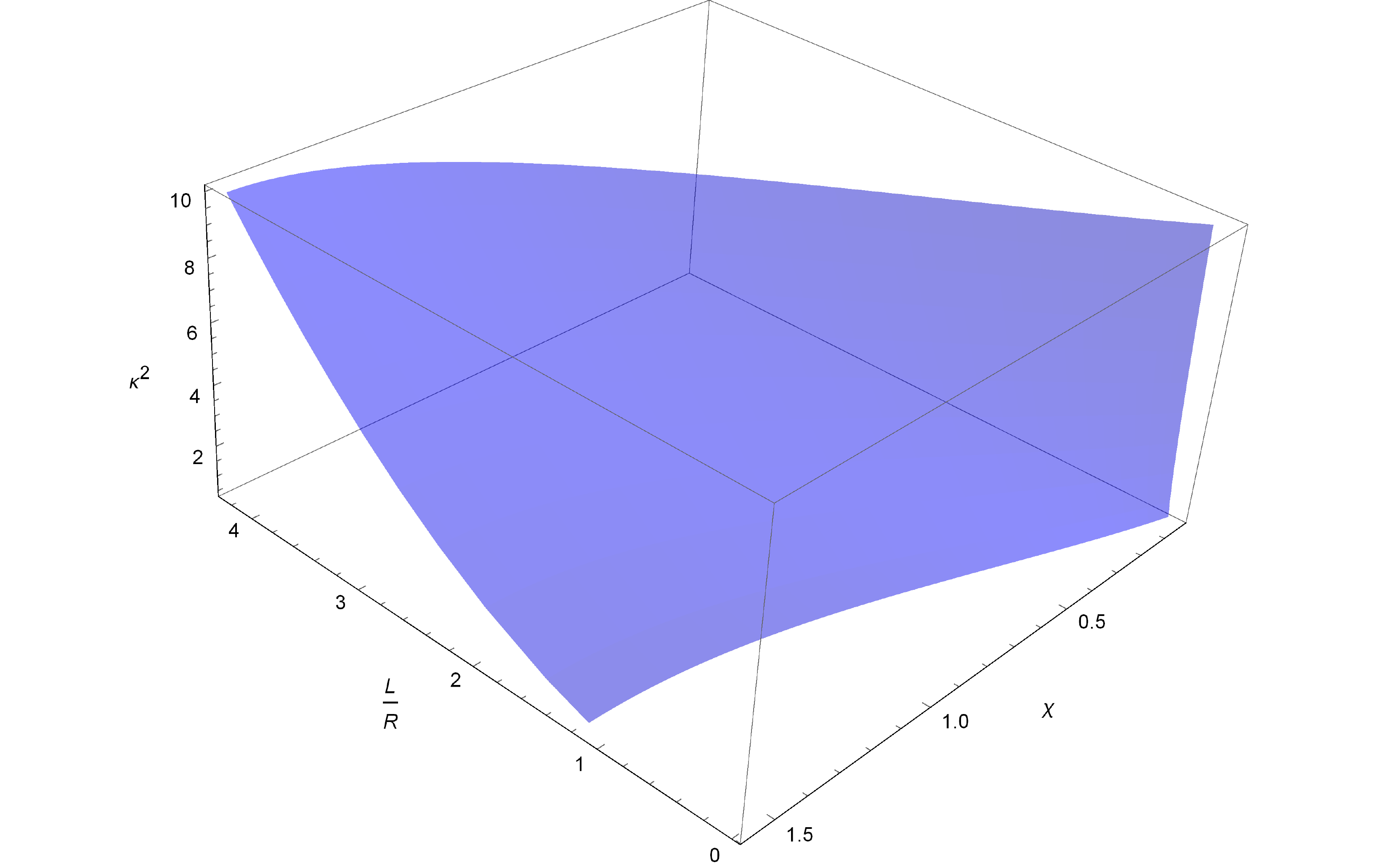}}
	\caption{\footnotesize \label{fig12} In the parameter space $(\chi,L/R,\chi)$ we have plotted the critical surface, namely the set of point for which the area of the connected solution is equal to the one of the dome.}
\end{figure}

In fig. \ref{fig12}, we have drawn the critical surface in the parameter space $(\chi,L/R,\chi)$  that corresponds to the locus of  the first order phase transition, choosing values of the fluxes up to ten and of the distances up to four. The surfaces when $\chi$ approaches to 0 collapses to a point again suggesting that there is no connected BPS solution.

\section{Perturbation Theory} 
\subsection{Perturbation theory: the non-BPS case}
\label{NBPS}
In order to obtain explicit results at perturbative level one should consider field configurations expanded around the supersymmetric vacuum \eqref{eq:Fuzzy funnel VEV}: the determination of the effective propagators and interaction vertices in this background needs a careful and non-trivial diagonalization procedure, that has been presented in \cite{Buhl-Mortensen:2016jqo}. At one-loop level the effect of the defect is entirely encoded into bulk propagators and vertices: the full Wilson loop expectation value is therefore given at this order by the tree-level and one-loop contributions
\begin{equation}
\left\langle {\cal W}\right\rangle=\left\langle {\cal W}\right\rangle _{\left(0\right)}+\left\langle {\cal W}\right\rangle _{\left(1\right)}
\end{equation}
The relevant computation has been already presented in \cite{Aguilera-Damia:2016bqv}: indeed, already at tree-level, they found
\begin{equation}
\left\langle {\cal W}\right\rangle _{\left(0\right)}=N-k+\frac{\sinh\left(\frac{\pi R\sin\chi}{L}k\right)}{\sinh\left(\frac{\pi R\sin\chi}{L}\right)},\label{eq:Correa tree level contribution}
\end{equation}
with $N-k$ corresponding to the standard tree-level contribution for the circular Wilson loop while the second term is the interesting one for the comparison with the connected string result. The one-loop part instead is given by
\begin{equation}
\left\langle {\cal W}\right\rangle _{\left(1\right)}=W_{\left(1\right)}+\frac{g_{YM}^{2}\left(N-k\right)R}{4\pi L}\int_{0}^{\infty}\!\!\!\!dr~r\!\int_{-\pi}^{\pi}\!\!\!d\delta\displaystyle\frac{\sinh\left(\frac{\left(\pi-\delta\right)R\sin\chi}{2L}k\right)}{\sinh\left(\frac{\left(\pi-\delta\right)R\sin\chi}{2L}\right)}\left(\mathcal{I}_{1}+\sin^{2}\chi\mathcal{I}_{2}\right),
\end{equation}
where $W_{\left(1\right)}$ contains the standard contribution for the non-broken
theory with $N\to(N-k)$ and another one scaling as $k^2$ (the two terms were named $T_1$ and $T_4$ in \cite{Aguilera-Damia:2016bqv}). Because we want to compare the string result with the perturbative computation, we can safely discard both of them, the latter being subleading with the others in the large-N limit while the former should come from string solutions that do not end on the D5-brane, not giving  $\lambda\over k^2$ dependent corrections. 
More explicitly we have \cite{Aguilera-Damia:2016bqv}
\begin{equation}
\mathcal{I}_{1}=2\cos\frac{\delta}{2}\sin\left(\frac{2Rr}{L}\cos\frac{\delta}{2}\right)I_{\frac{k}{2}}\left(r\right)K_{\frac{k}{2}}\left(r\right),
\end{equation}
\begin{equation}
\mathcal{I}_{2}=\frac{\sin\left(\frac{2Rr}{L}\cos\frac{\delta}{2}\right)}{\cos\frac{\delta}{2}}\left(\frac{k-1}{2k}I_{\frac{k+2}{2}}\left(r\right)K_{\frac{k+2}{2}}\left(r\right)+\frac{k+1}{2k}I_{\frac{k-2}{2}}\left(r\right)K_{\frac{k-2}{2}}\left(r\right)-I_{\frac{k}{2}}\left(r\right)K_{\frac{k}{2}}\left(r\right)\right).
\end{equation}
As expected these expressions depend on $R$ and $L$ only through the ratio $\frac{R}{L}$. When $\chi\neq 0$ the integrals are very difficult to be performed analytically but, in the limit of large $k$, they can evaluated to extract their $\lambda/k^2$ behavior. The relevant technique has been already settled in \cite{Aguilera-Damia:2016bqv}, where the focus was on the limit $L/R\to 0$. As far as $\sin\chi\neq 0$ both $\left\langle {\cal W}\right\rangle _{\left(0\right)}$ and $\left\langle {\cal W}\right\rangle _{\left(1\right)}$ exhibit an exponential behavior in $k$, that can be evaluated elementary in $\left\langle {\cal W}\right\rangle _{\left(0\right)}$ and through a saddle-point approximation in the integrals defining $\left\langle {\cal W}\right\rangle _{\left(1\right)}$. We stress that the ratio $R/L$ can be taken finite in this computation, sending just the parameter $k$ to infinity\footnote{Of course in so doing we get also other terms, coming from the $\sinh$ factors at denominator in the relevant integrals. It is easy to realize that they cannot be produced from the classical connected string solution: we argue that they could be obtained from the one-loop corrections at strong coupling.}. Repeating the same same step as in \cite{Aguilera-Damia:2016bqv} we end up with
\begin{equation}
\log\left\langle {\cal W}\right\rangle\simeq-\frac{k\pi R}{L}\left(\sin\chi+\frac{\lambda}{4\pi^{2}k^{2}}\left(\frac{\pi}{2}-\chi-\frac{1}{2}\sin2\chi\right)\sec^3\chi \left(\sin^{2}\chi+\left(\frac{L}{R}\right)^{2}\right)+O\left( \frac{\lambda^2}{\pi^4k^4}\right) \right).\label{eq:log<W>}
\end{equation}
We will see in the next section that a perfect match with the strong coupling result at order $\frac{\lambda}{k^2}$ is obtained, without resorting to any limit on $R/L$.
\subsection{Comparing perturbative analysis with the strong coupling analysis}
\label{subsect:strongcouplingexpansion}
Usually, in the AdS/CFT correspondence, perturbative computations and the supergravity analysis live in opposite regime and cannot be successfully compared. For this class of defect conformal field theories, one can instead consider a double-scaling limit
\cite{Nagasaki:2011ue,deLeeuw:2016vgp}  that opens a new window.
Gravity computations, which are valid for large 't Hooft coupling 
$\lambda$, can be considered for large $k$ in such a way that  $\lambda/k^2$ is kept small and the results are found to be expressible in powers of $\lambda/k^2$. Thus, in this regime, it is possible to successfully compare gauge and gravity results providing further non-trivial verifications of the AdS/CFT correspondence.

In the strong coupling regime, this limit is equivalent to expand our classical solution in power of $\frac{1}{\kappa^2}$, namely for large value of the flux. The flux diverges when the denominator in eq. \eqref{eq:2.36} vanishes, but the numerator does not. This occurs when $x$,  $m$ and $\chi$ satisfy the relation
\be
\label{eq:4.1}
m~ x^2~
   \text{cn}\left(\left.\frac{1}{2} x (\pi -2 \chi )\right|m\right)^2+x^2-1=0.
\ee
Even though we are considering the regime  in which $\kappa^2$ is very large and eventually diverges, we require that the (adimensional) distance $L/R$ of the Wilson loop  from the defect remains finite.  The 
parameter $\eta=\text{arcsinh}\frac{L}{R}$ in terms of $x$, $m$ and $\chi$ is the sum of  two positive contributions. In fact,  in eq.   \eqref{eq:2.43} the first term is the integral of a positive function, while the second one is the $\tanh ^{-1}$ of a positive argument.
If use the eq. \eqref{eq:2.36} we can recast the argument of $\tanh ^{-1}$  in the following form:
\be
\label{eq:4.2}
 \sqrt{\frac{2\kappa ^2 \left(1-(m+1) x^2\right)}{\kappa ^2 \left(1-(m+1) x^2\right)-(m+1) x^2+\sqrt{\left(\kappa ^2 \left(1-(m+1) x^2\right)-(m+1) x^2\right)^2-4 \left(\kappa ^2+1\right) m x^4}+2}},
\ee
which allows us to study easily its behavior for large $\kappa^2$. If the $m$ and $x$ are finite when $\kappa^2\to\infty$ and the combination $1-(m+1) x^2$ does not vanish, it is  straightforward to realize that the quantity in  eq. \eqref{eq:4.2} approaches $1$. Therefore 
its contribution to the distance ($\tanh ^{-1}(1)$) diverges. To avoid this conclusion we must require 
\be
\label{eq:4.3}
x=\frac{1}{\sqrt{1+m}}+O\left(\frac{1}{\kappa^2}\right)\ \ \ \ \ \ \ \kappa\to\infty.
\ee
If we use the definition of the unknown $x$,   the above result is equivalent to require that  $j^2=O(\kappa^2)$, therefore we are in the region (A).  In this limit, with the help of  the expansion eq. \eqref{eq:4.3}, eq. \eqref{eq:4.1} collapses to
\be
\label{eq:4.4}
\frac{m}{m+1}~ 
   \text{sn}\left(\left.\frac{1}{2\sqrt{1+m}} x (\pi -2 \chi )\right|m\right)^2=0\ \ \ \Rightarrow \ \ \ \ m=0.
\ee
The requirement that the distance is kept finite and different from zero fixes as  $m$ must vanish for large $\kappa$: $m=O\left(\frac{1}{\kappa^2}\right)$.  Therefore, in this regime  we can safely assume the following expansion for the  parameters:
\be
m=\sum_{n=1}^{\infty}  \frac{a_{2n}}{\kappa^{2n}}\ \ \ \ \ \ \ \ \ \ \ \ \ \ \ \ \ \ \ \ \  x=1+\sum_{n=1}^{\infty}  \frac{b_{2n}}{\kappa^{2n}}
\ee
The coefficients can be determined by solving iteratively the equation for the flux and the condition fixing the distance (see app.  \ref{app:largekespansion}). At the  lowest  order we get 
\begin{subequations}
\begin{align}
m=&\frac{\sec ^2\chi }{\kappa^2}\left(1-\left(\frac{L^2}{R^2}+1\right) \sec ^2\chi \right)+O\left(\frac{1}{\kappa^4}\right)\\
x=&1+\frac{1}{2\kappa^2} \left(\frac{L^2}{R^2}+1\right) \tan ^2\chi  \sec ^2\chi +O\left(\frac{1}{\kappa^4}\right)
\end{align}
The subsequent terms in the expansion are quite cumbersome and their explicit form up to $1/\kappa^4$  is given in app. \ref{app:largekespansion}.
\end{subequations}
We can now use this expansion (see app. \ref{app:largekespansion}) to determine the first two terms in the large $\kappa$ expansion
\be
\begin{split}
{S}=&-\frac{\sqrt{\lambda}\kappa R}{L}\left[
 \sin \chi+\frac{1}{4 \kappa^2 
   }{ \left(\frac{\pi}{2}-\chi -\frac{1}{2}\sin 2\chi \right) \sec ^3\chi  \left( \sin^2\chi+\frac{ L^2}{R^2}\right)}\right]+O\left({\kappa }^{-2}\right)=\\
 =&  -\frac{\pi k R}{L}\left[
 \sin \chi+\frac{\lambda}{4 \pi^2 k^2 
   }{ \left(\frac{\pi}{2}-\chi -\frac{1}{2}\sin 2\chi \right) \sec ^3\chi  \left( \sin^2\chi+\frac{ L^2}{R^2}\right)}+O\left(\frac{\lambda^2}{\pi^4 k^4}\right)\right],
   \end{split}
\ee
where we have replaced the strong coupling quantity $\kappa$ with its expression in terms of $\sqrt{\lambda}$ and the integer flux $k$. 
A simple power-counting of the coupling constant immediately shows that this are only two terms which can be compared with our perturbative computation and they
successfully reproduce eq. \eqref{eq:log<W>}.

\subsection{Perturbation theory: the BPS case}
\label{perturbBPS}
We have already seen that at $\chi=0$, when the Wilson loop only couples to the massless scalar $\Phi_6$ and we are therefore at the BPS point, the connected string solution does not exist. At weak coupling we expect conversely that the scaling $\lambda/k^2$ should break: let us examine more closely the situation. We start by observing that a dramatic simplification occurs at the perturbative level, the non-trivial contribution at one-loop reducing to a much simpler expression
\begin{equation}
\left\langle {\cal W}\right\rangle _{\left(1\right)}=W_{\left(1\right)}+\frac{g_{YM}^{2}\left(N-k\right)Rk}{ 4\pi L} \int_{0}^{\infty}\!\!\!\!dr~r\!\int_{-\pi}^{\pi} d\delta~ ~ 2\cos\frac{\delta}{2}\sin\left(\frac{2Rr}{L}\cos\frac{\delta}{2}\right)I_{\frac{k}{2}}\left(r\right)K_{\frac{k}{2}}\left(r\right)\,.
\end{equation}
The integral over the angular variable $\delta$  can be exactly performed in terms of the Bessel function $J_1$ leading to
\begin{equation}
\left\langle {\cal W}\right\rangle _{\left(1\right)}=W_{\left(1\right)}+\frac{g_{YM}^{2}\left(N-k\right)Rk}{L}\int_{0}^{\infty}\!\!\!\!dr\,r\, J_1\left(\frac{2Rr}{L}\right)I_{\frac{k}{2}}\left(r\right)K_{\frac{k}{2}}\left(r\right)\,,
\end{equation}
and the radial integral is solved in terms of Meijer G Functions:
\begin{equation}
\begin{split}
\left\langle {\cal W}\right\rangle _{\left(1\right)}&=
W_{\left(1\right)}+\frac{g_{YM}^{2}\left(N-k\right)k L}{ 4 \sqrt{\pi }R}\,G_{3,3}^{2,2}\left(\frac{R^2}{L^2}\Biggl|
\begin{array}{c}
1,1-\frac{k}{2},\frac{k+2}{2} \\
\frac{1}{2},\frac{3}{2},\frac{1}{2} \\
\end{array}
\right)\,.
\end{split}
\end{equation}
We shall explore two interesting limits of this exact result. First we investigate the behavior of  $\left\langle {\cal W}\right\rangle _{\left(1\right)}$  when the parameter $k$ goes to $\infty$. This limit is easier to discuss if we step back to the integral form of the one-loop expression and perform the change of variables  $r\to \frac{k}{2} r$
\begin{equation}
\left\langle {\cal W}\right\rangle _{\left(1\right)}=W_{\left(1\right)}+\frac{g_{YM}^{2}\left(N-k\right)R}{4 L} k^3\int_{0}^{\infty}\!\!\!\!dr\,r\, J_1\left(\frac{R k r}{L}\right)I_{\frac{k}{2}}\left(\frac{k r}{2}\right)K_{\frac{k}{2}}\left(\frac{k r}{2}\right).
\end{equation}
Then we can use the following asymptotic expansions for the product of the modified Bessel functions
\be
K_{\frac{k}{2}}\left(\frac{k r}{2}\right)
I_{\frac{k}{2}}\left(\frac{k r}{2}\right)=\frac{1}{ k (1+ r^2)^{1/2}}\sum_{n=0}^\infty \frac{2^{2n}}{k^{2n}}\sum^{2n}_{m=0} (-1)^m U_{2n-m}( (1+ r^2)^{-\frac{1}{2}})U_m( (1+ r^2)^{-\frac{1}{2}}),
\ee
which is obtained by combining the expansion 10.41.3 and 10.41.4 of \cite{Nist}. The  $U_n(x)$ are polynomials that can be constructed recursively (see sec. 10.41 in \cite{Nist} for details): $U_0(x)=1, U_1(x)=\frac{x}{8}-\frac{5 x^3}{24},\dots$. If we are interested at the leading order in $k$, it is sufficient to consider the first term in the expansion:
\begin{equation}
\label{WL22}
\begin{split}
\left\langle {\cal W}\right\rangle _{\left(1\right)}&\simeq W_{\left(1\right)}+\frac{g_{YM}^{2}\left(N-k\right)R}{4 L}  k^2\int_{0}^{\infty}\!\!\!\!dr J_1\left(\frac{R k r}{L}\right)\frac{r}{  (1+ r^2)^{1/2}}=\\
=&W_{\left(1\right)}+\frac{g_{YM}^{2}\left(N-k\right)L}{4 R}  \frac{2}{\sqrt{\pi }  } G_{1,3}^{2,1}\left(\frac{k^2 R^2}{4 L^2}\Biggl|
\begin{array}{c}
1 \\
\frac{1}{2},\frac{3}{2},\frac{1}{2} \\
\end{array}
\right)=\\
=&W_{\left(1\right)}+\lambda\frac{L}{4 R} \left(1+O\left(\frac{1}{k^2}\right)\right).
\end{split}
\end{equation}
The BPS circle does not possess the correct scaling to match a potential connected string solution: its expansion does not organize in a $\lambda/k^2$ series. This corroborates the absence of a connected solution the BPS case. \vspace{0.3cm} \\
The second limit we shall consider is $R\to 0$, namely when we shrink the loop to a point. Since we are dealing with a conformal field theory, it is the dimensionless combination $x=R/L$,  which approaches to zero: this should also correspond to place the Wilson loop at infinite distance from the defect, recovering at leading order the expectation value of the usual BPS circle. We find
\begin{equation}
\label{WL21}
\begin{split}
\left\langle {\cal W}\right\rangle _{\left(1\right)}&=
W_{\left(1\right)}\!+\!\frac{ g_{YM}^2 k (N-k)}{4}\left[1\!-\!\frac{1}{4}   \left(1\!-\!k^2\right)
x^2  \left(\psi\left(\frac{k+1}{2}\right)\!+\!\gamma_E+ \log x\!-\!\log 2\right)\!-\!\right.\\
& \hspace{5cm}-\left.\frac{x^2}{2}+O\left(x^4\right)\right].
\end{split}
\end{equation}
As expected we see that the leading term scales as a constant in this limit, and when combined with $W_{(1)}$ it reconstructs at this order the circular Wilson loop in absence of defect. Actually the next term in the expansion can be easily understood in terms of the operator product expansion (OPE) of the circular BPS loop. 
\subsection{Operator product expansion expansion of the Wilson loop}
\label{OPE1}
In absence of defect the Wilson loop, when probed from a distance much larger than the size of the loop itself, can approximated in CFT by an expansion of local operators \cite{Berenstein:1998ij,Arutyunov:2001hs}
\be
\label{eq:2.5.1}
\frac{\mathcal{W}(\mathcal{C})}{\left\langle    \mathcal{W}(\mathcal{C}) \right\rangle}= \mathds{1}+\sum_{k}c_k R^{\Delta_k}\,\mathcal{O}^{(k)} (x)\,,
\ee
where $R$ is the radius of the loop, the $\mathcal{O}^{(k)}$ are composite operators with conformal weights $\Delta_{k}$ evaluated at the center of the loop, and $c_{k}$ are the OPE coefficients that depend on $\lambda$. In perturbation theory the scaling dimension of an operator can be represented as
\be
\label{eq:2.5.2}
\Delta=\Delta^{(0)}+\Delta^{(1)}+\Delta^{(2)}+\cdots\,,
\ee
where $\Delta^{(0)}$ is the free field dimension, and $\Delta^{(1)},\Delta^{(2)}$ are anomalous dimensions at order $\lambda$, $\lambda^2$ and so on. Since the symmetries of a CFT constraint the one-point functions of operators that are not the identity to be zero in absence of defects, the expectation value of the Wilson loop corresponds to the coefficient of the identity. The OPE coefficients can be computed perturbatively, but the result cannot be extrapolated in general to strong coupling. The operators appearing in eq. \eqref{eq:2.5.1} must have the same properties as the Wilson loop, the $\mathcal{O}^{(k)}$ being therefore bosonic and gauge invariant. The possible contributions to the OPE for low value of the scaling dimension are
\begin{itemize}
	\item $\Delta^{(0)}=0:$ the only possible contribution comes from operators proportional to the identity;
	\item $\Delta^{(0)}=1:$ the only elementary fields with scaling dimension one are the scalars. The trace of a single scalar is the only gauge-invariant operator, but since the $\Phi_I$'s are valued in the Lie algebra  $\mathfrak{su}(N)$ this contribution vanishes;
	\item  $\Delta^{(0)}=2:$ the only two types of gauge-invariant operators are the chiral primary operators $\mathcal{O}^a$ and the Konishi scalar $\mathcal{K}$. They are canonically normalized as
	\be
	\label{eq:2.5.3}
	\mathcal{O}^a=4\sqrt{2}\p^2C^{a}_{IJ}\,\text:\text{Tr}(\Phi^I\Phi^J):\,, \qquad \mathcal{K}=\frac{4\p^2}{\sqrt{3}}:\text{Tr}(\Phi^I\Phi^I):.
	\ee
	Here the traceless symmetric tensor $C^a_{IJ}$ obeys $C^a_{IJ} C^b_{IJ}=\delta^{ab}$ with $a,b=1,\dots,20.$ The operators $\mathcal{O}^a$ lie in a short supermultiplet and transform in the $\mathbf{20}$ irreducible representation of the R-symmetry group $SO(6)_R$. They have vanishing anomalous dimension. The Konishi scalar is the lowest component of the long supermultiplet \cite{Konishi:1983hf}, and it acquires an anomalous dimension in perturbation theory. Its one-loop anomalous dimension is $\Delta^{(1)}=\frac{3\lambda}{4\p^2}$ \cite{Anselmi:1998ms,Bianchi:1999ge,Arutyunov:2000ku,Arutyunov:2000im,Arutyunov:2001mh,Penati:2001sv}.
\end{itemize}
Given the expansion eq. \eqref{eq:2.5.2} for the scaling dimension, every term $R^{\Delta}$ in eq. \eqref{eq:2.5.1} produces logarithmic terms as
\be
\label{eq:2.5.5}
R^{\Delta}=R^{\Delta^{(0)}}\left(1+\Delta^{(1)}\log R+\Delta^{(2)}\log R+\frac{1}{2}\Delta^{(1)^2}\log^2 R+\cdots\right) \,,
\ee
 in the limit $R \rightarrow 0$ the coefficients of the divergent logarithms $\log R$ is proportional to the one-loop anomalous dimensions of the non-protected operators appearing in the Wilson loop OPE, as the Konishi operator. The OPE coefficients can be read off from the correlation functions of the Wilson loop with local operators: in particular one can consider CPOs with scaling dimension $\Delta\equiv k$ defined as
   \be
   \label{eq:2.5.6}
   \mathcal{O}_{k}^I(x)=\frac{(8\p^2)^{k/2}}{\sqrt{k}}C^I_{J_1\cdots J_{k}}:\text{Tr}(\Phi^{J_1}\cdots\Phi^{J_{k}} ) :\,,
   \ee
   where $C^I_{J_1\cdots J_k}$ are totally symmetric traceless tensors normalized as $C^I_{J_1\dots J_k}C^L_{J_1\dots J_k}=\delta^{IL}$. If $k=2$, one obtains the chiral primary operator $\mathcal{O}^a$ with $\Delta=2$.  It is possible to show that their two-point functions are protected by supersymmetry and their scaling dimensions do not receive radiative corrections. From the exact expression for correlators of the circular Wilson loop with CPOs eq. \eqref{eq:2.5.6} valid for any $\lambda$ found in \cite{Semenoff:2001xp} it is possible to recover the relevant OPE coefficients at any order
   \be
   \label{eq:2.5.8}
   c^I_{k}=\frac{2^{k/2-1}}{N}\sqrt{\frac{k}{ \lambda}}\,\frac{I_k(\sqrt{\lambda})}{I_1(\sqrt{\lambda})}Y^{I}(\theta),
   \ee
   where $I_{k}$ and $I_1$ are modified Bessel functions and $Y^I(\theta)$ are spherical harmonics
   \be
   Y^I(\theta)=C_{J_1\dots J_k}^I\theta^{J_1}\cdots\theta^{J_k}
   \ee
  with the index $I$ running over all the spherical harmonics of $SO(6)$ Casimir \cite{Lee:1998bxa}. Perturbatively, the leading contribution to the correlation functions of the circular Wilson loop with the chiral primary with smallest conformal dimension and the Konishi operator was found in \cite{Arutyunov:2001hs} giving
\be
c^a_\mathcal{O}=\frac{1}{2\sqrt{2}N}Y^a(\theta)\,, \qquad c_1=\frac{1}{4\sqrt{3}N}.
\ee
Summarizing the lower dimensional content of the local operator expansion, we have 
\be
\begin{split}
\label{eq:2.5.13}
\frac{\mathcal{W}(\mathcal{C})}{\left\langle \mathcal{W}(\mathcal{C})\right\rangle }=& \mathds{1}+R^{\Delta_{\mathcal{K}}}\left(\frac{1}{4\sqrt{3}N}+\frac{\lambda c_2}{N} +\cdots\right)\mathcal{K}(x)+\\ & +R^{\Delta_\mathcal{O}}\left(\frac{1}{2\sqrt{2}N} -\frac{\lambda}{48\sqrt{2}N}+ \cdots\right)Y_{a}(\theta)\mathcal{O}^a(x)+\text{higher scaling dimension}\,,
\end{split}
\ee
the dots indicate higher order terms in $\lambda$ of the corresponding operator expansion coefficients. The value of $c_2$ for the Konishi operator has not been computed, at least at our knowledge, while the OPE coefficient at order $\lambda$ for the chiral primary operator with $k=2$ is obtained expanding for small $\lambda$ the r.h.s of eq. \eqref{eq:2.5.8}. The scaling dimension of $\mathcal{O}^a$ is $\Delta_{\mathcal{O}} = 2$ and this operator does not get an anomalous dimension, whereas $\Delta_{\mathcal{K}}$ receives perturbative corrections and following eq. \eqref{eq:2.5.5} one can write
\be
R^{\Delta_{\mathcal{K}}}=R^2\left(1+\Delta_{\mathcal{K}}^{(1)}\log R +\cdots\right) \,.
\ee
\subsection{Wilson loop OPE and one-point functions}
\label{OPE}
In the presence of a defect we expect that the structure of the OPE for the circular Wilson loop is unchanged, due to the fact that we are effectively probing the operator at infinite distance or, alternatively, because the ultraviolet properties of the theory are insensible to the boundary. The only modification needed is implied by the presence of non-trivial one-point functions: the Wilson coefficient of the identity is unchanged and takes into account still the contribution of the expectation value ${\langle \mathcal{W}\rangle_0 }$ in absence of defect. We are led therefore to assume the following OPE in the defect theory
\be
\begin{split}
\label{WE}
\frac{\mathcal{W}}{\langle \mathcal{W}\rangle_0 }=& \mathds{1}+R^{\Delta_{\mathcal{K}}}\left(\frac{1}{4\sqrt{3}N}+\frac{\lambda c_2}{N} +\cdots\right)\mathcal{K}(x)+\\ & +R^{2}\left(\frac{1}{2\sqrt{2}N} -\frac{\lambda}{48\sqrt{2}N}+ \cdots\right)Y_{a}(\theta)\mathcal{O}^a(x)+\text{higher scaling dimension}\,,
\end{split}
\ee
the expansion being normalized using $\langle \mathcal{W}\rangle_0$.  For the explicit definition of $C^a_{ij}$ and $Y^a$ see Appendix \ref{Basis}. Now taking the vacuum expectation value of eq. \eqref{WE} we understand that the expansion we have derived for the Wilson loop for $R/L\to 0$ (see eq. \eqref{WL21}) should be recovered from the one-point functions of the Konishi operator $\mathcal{K}(x)$ and of the combination of chiral primaries $Y_{a}(\theta)\mathcal{O}^a(x)$, from the one-loop anomalous dimension of $\mathcal{K}(x)$ and from the Wilson coefficient $c_2$. Fortunately in a beautiful series of papers \cite{deLeeuw:2015hxa,Buhl-Mortensen:2015gfd,Buhl-Mortensen:2017ind} the NBI group has studied the one-point functions of scalar operators in the defect theory, obtaining explicit result both at tree and one-loop level through perturbative computations and at all-order applying integrability techniques. We take advantage of their efforts and we adapt their results to our relevant operators (see Appendix \ref{Basis} for the full details)
\be
\langle Y_{a}(\theta)\mathcal{O}^a(x)\rangle=\frac{\pi^2}{3\sqrt{2} L^2} k(1-k^2)- \frac{\lambda }{2\sqrt{2} N L^2} \left(k(N-k)+\frac{k^2-1}{2}\right)
\ee
\be
\langle\mathcal{K}(x)\rangle=-\frac{\pi^2}{\sqrt{3} L^2} k(1-k^2)-\frac{\sqrt{3}\lambda }{4 L^2}   k\left(1-k^2\right) \left(\psi\left(\frac{k+1}{2}\right)+\gamma_E-\log 2+\frac{5}{6}\right).
\ee
Using further the one-loop contribution to the anomalous dimension of the  Konishi operator $ \Delta^{(1)}=\frac{3\lambda}{4\pi^2} $ we can compare the OPE expansion with the direct computation of the Wilson loop in the small $R/L$ limit eq. (\ref{WL21}). We see that our result non-trivially matches with the one-point functions derived in \cite{deLeeuw:2015hxa,Buhl-Mortensen:2015gfd,Buhl-Mortensen:2017ind} if
\be
c_2=-\frac{15+\p^2}{96 \sqrt{3}\p^2}
\ee
Thus, from the OPE for the Wilson loop in the defect case, we have a prediction for the Wilson coefficient of the Konishi operator at order $\lambda$ in eq. \eqref{eq:2.5.13}. This prediction could be verified by computing the two-loop contribution to the two-point function of the Konishi operator with the circular Wilson loop in absence of defect.
\section{Conclusions and outlook}
\label{conclusions}
The introduction of defects in conformal field theories implies, in general, an augment of the independent conformal data and enriches the dynamics with novel effects that certainly deserve further studies. While at level of correlation functions of local operators there has been a considerable amount of investigations in this field, much less attention has been devoted to the behavior of non-local operator: in this paper we tried to fill partially this gap, studying the fate of the circular Wilson loop in the defect ${\cal N}=4$ Super Yang-Mills theory both at strong and weak coupling. In the former case, using AdS/CFT correspondence, we have explored in full generality the structure of the vacuum expectation value of a loop parallel to the defect, finding the semiclassical string solution in the complete parameter space and computing the related classical action. The main result has been the discovery of a novel Gross-Ooguri type transition, separating a phase in which the dome solution, associated with the Wilson loop in absence of defect, dominates from a situation in which a cylindrical minimal surface attached to the defect D5 brane describes the non-local operator. In the generic case, we have performed a double-scaling limit on our cylindrical solution, sending $k\to\infty$ with $\lambda/k^2$ fixed, recovering without resorting to any geometrical approximation the perturbative Feynman diagram result. For the particular case in which the Wilson loop operator becomes BPS, $i.e.$ for 
$\chi=0$ in our notation, the connected solution ceases to exist and the strong coupling regime is arguably described by supergravity exchanges between the spherical dome and the D5 brane. Conversely, we found an analogous behavior at weak coupling, the BPS case not respecting the expected double-scaling limit. For the BPS case we have also explored at one-loop the shrinking (or equivalently the large distance) behavior, finding that it can be nicely understood in terms of the OPE of the Wilson loop operator: the knowledge of the non-trivial one-point functions for scalar operators of classical dimension two allows to reconstruct explicitly the first terms of the expansion. Assuming a certain value for the one-loop contribution to the relevant Wilson coefficient of the Konishi operator we find a perfect matching between our computations and the results of 
\cite{deLeeuw:2015hxa,Buhl-Mortensen:2015gfd,Buhl-Mortensen:2017ind}. 

There are a number of different directions that can be explored in order to improve the present investigations. First of all one should compute independently the Wilson coefficient for the Konishi in the circular BPS Wilson loop OPE: this would represent a non-trivial check of the result obtained for one-point functions in \cite{deLeeuw:2015hxa,Buhl-Mortensen:2015gfd,Buhl-Mortensen:2017ind} or would enlighten potential subtleties in our OPE description. A second and intriguing question concerns the short-distance limit from the defect (or equivalently the large radius limit of the loop) in the BPS case. Let us consider the expansion of eq. \eqref{WL22} in the limit $L\to 0$ namely $x\to \infty$: for odd values of $k$ we find an analytical series in inverse odd powers of $x$: 
\be
 \left\langle {\cal W}\right\rangle _{\left(1\right)}=  W_{\left(1\right)}+\frac{g_{YM}^{2}\left(N-k\right) }{ 4 }\left[  \frac{1}{ x}-\frac{3}{\left(8 -2 k^2\right)
   x^3}+O\left(\left(\frac{1}{x}\right)^4\right)
   \right].
   \ee 
For even $k$ instead we observe the appearance also of logarithmic corrections that start at the order $x^{-k-1}$. For instance for $k=4$ we find
 \be
   W_{\left(1\right)}+g_{YM}^{2}\left(N-4\right)  \left[
 \frac{1}{4 x}+\frac{1}{32 x^3}+\frac{3 (120 \log
   (8x)-289)}{8192 x^5}-\frac{25 (168 \log (8x)-367)}{65536 x^7}+O\left(\frac{1}{x^8}\right)\right].
 \ee
It would be tempting to interpret these expressions in terms of a boundary operator expansion (BOE) of the Wilson loop: BOE in defect ${\cal N}=4$ SYM has been already considered for scalar two-point functions in \cite{deLeeuw:2017dkd}. In this paper the spectrum of gauge-invariant boundary operators of the theory has been also presented (see also \cite{Ipsen:2019jne}): it would be interesting to derive a version of the BOE for the circular BPS Wilson loop and to use the consistency of bulk and boundary operator expansions to get new information on the defect theory. A puzzling aspect of the above computations is that their analytical properties depend crucially on $k$. For odd $k$ the absence of logarithm suggests that only protected operators should appear in the BOE, while for even $k$ also non-protected operators seem to be part of the game.

In the BPS case we have found an exact analytical expression for the vacuum expectation value of the Wilson loop at the first perturbative order and we have observed that no connected string solution appears at strong coupling: these two facts might signal that an exact evaluation of this Wilson loop could be feasible, resorting probably to a highly non-trivial application of supersymmetric localization in this context.

A more direct and conceptually straightforward follow-up of our investigations concerns the case of the correlator of two circular Wilson loop in the defect set-up \cite{avvenire}: the case of two straight-line has been already tackled in \cite{Preti:2017fhw} where a complicate pattern of Gross-Ooguri like phase transition has been discovered for the quark-antiquark potential. In the circular case, the situation is more complicated because of the larger parameter space and the possibility to have both "undefected" connected string solutions between the two circles or individual cylinder/dome solutions dominating the semiclassical strong coupling regime. Non-trivial string three-point functions could also enter the game, describing new connected minimal surfaces with three holes, one of which lying on the D5 brane.

Finally, the generalization of our investigation in the cousin theories related to the non-supersymmetric D3/D7 system could be certainly considered.

\subsection*{Acknowledgements}
Special thanks go to Marisa Bonini for participating at the early stages of this work and many interesting discussions. It is also our pleasure to thank L. Bianchi, A. Bissi, N. Drukker, C. Kristjansen, S. Penati, M. Preti, D. Trancanelli, E. Vescovi and  K. Zarembo for useful discussions. 
\newpage

 \appendix
\section{Expression for the distance from the defect}
\label{distance}
The distance  eq. \eqref{eq:2.43}  from the defect brane can be  also expressed in terms of the elliptic integral of the third kind if we explicitly 
perform the integration defining the auxiliary function $v(\sigma)$.  We obtain
\begin{align}
\label{eq:3.1}
 \eta=&\sqrt{-\frac{(j^2 m+1)(j^2+m)}{(m+1)(j^2-1)}}\left[\sqrt{\frac{j^2-1}{m+1}} \tilde\sigma- \Pi \left(-\frac{m+1}{j^2-1};\left.\text{am}\left(\left.\sqrt{\frac{j^2-1}{m+1}} \tilde\sigma\right|m\right)\right|m\right)\right]+\noindent\nonumber\\
&\ \ \ \ \ \ \ \ \ \ \ \ \ \ +\tanh ^{-1}\!\!\left(-\sqrt{-\frac{(m+1)^2}{(j^2 m+1)(j^2+m)}}
\frac{g^\prime(\tilde\sigma )}{g(\tilde\sigma
   )}\right)=\\
   =&\sqrt{-\frac{(n+1)(n m+1)}{n}}\left[\sqrt{n}\tilde\sigma-\Pi\left(\left.-\frac{1}{n}; \text{am}(\sqrt{n}\tilde\sigma|m)\right|m\right)\right]\!+\!
   \tanh ^{-1}\!\!\left(\frac{-\frac{g^\prime(\tilde\sigma )}{g(\tilde\sigma
   )}}{\sqrt{-(n m+1)(n+1)}}\nonumber
\right)
\end{align}
   where we have introduced the short-hand notation 
   \be
   n\equiv\frac{j^2-1}{m+1}.
   \ee
This second representation of the distance will be useful when doing analytical expansions, the one provided by eq.   \eqref{eq:2.43} being more suitable in numerical analysis. 

For instance, from eq. \eqref{eq:3.1} is quite straightforward to see that the distance vanishes  when  $m$ approaches $m_c$. We first observe that the argument of the  $\text{arctanh}$  behaves like
 \be
 \sim\frac{\sqrt{n}}{n \sqrt{-m_c}}\frac{\text{cn}(\sqrt{n} \tilde\sigma|m)\text{dn}(\sqrt{n} \tilde\sigma|m)}{\text{sn}(\sqrt{n} \tilde\sigma|m)}=
 \frac{s_0}{\sqrt{n} },
 \ee
 where we have taken into account that the combination $\sqrt{n} \tilde\sigma$ is finite in this limit: it vanishes since $n\to\infty$ as  
 $m \to m_c$ (see eq. \eqref{nmc}).The remaining  contribution  can be also seen to vanish
when we exploit the  behavior of the incomplete elliptic integral of   the third kind  for small values of its first argument. In fact we have
\cite{wolfram}
\be
\Pi\left(\left.-\frac{1}{n}; \text{am}(\sqrt{n}\tilde\sigma|m)\right|m\right)\simeq z+ {c(z,m)}\frac{1}{n} +O\left( \frac{1}{n^2}\right),
\ee
which in turn implies that this contribution also vanishes as $1/\sqrt{n}$.

\section{Expansion of  \texorpdfstring{$n$,  $g(\tilde\sigma)$}{n,g(s)} and \texorpdfstring{$\frac{\partial\eta}{\partial m}$}{eta(m)} close to the boundary \texorpdfstring{$j^2=-1/m^2$}{j2eqm1divm2}}
\label{Expansionm0}
In the parameter space the  boundary $j^2=-1/m^2$ corresponds to the curved red line plotted in fig. \ref{fig1a}. The values of $\chi$ and $m$ along this curve are related by eq.  \eqref{chim0}. Our goal is now to expand some relevant quantities in the region close to this boundary.  To begin with,  we shall choose a value for the angle: $\chi=\chi_0$.  Eq. \eqref{chim0} allows us to translate it in a value $m_0$ for $m$. Given the pair  $(\chi_0, m_0)$ on the red curve, the value  $x$ is fixed by eq. \eqref{eq:2.49a} to be
\be
x_0=1.
\ee	
We now expand around this configuration. Specifically we keep the angle $\chi_0$ fixed  and  we allow $m$ to be different from $m_0$, but close 
to it. Then  we can write $x$ as series expansion around $m_0$:
\be
x=1+s_1 (m-m_0)+s_2(m-m_0)^2+O((m-m_0)^3).
\ee
The coefficients $s_i$ can be determined by solving perturbatively eq. \eqref{eq:2.36}. With help of Mathematica and after some trivial manipulations one finds
\be
\label{expx}
x=1+\frac{\left(m-m_0\right){}^2 \left(\kappa ^2 m_0+m_0-1\right) \left(\left(m_0-1\right)
   \mathds{K}\left(m_0\right)+\mathds{E}\left(m_0\right)\right){}^2}{8 \kappa ^2 \left(m_0-1\right)
   m_0^2}+O\left(\left(m-m_0\right){}^3\right).
\ee
 Given eq.  \eqref{expx} the corresponding expansion for $n$ around $m=m_0$ is  easily recovered
\be
\begin{split}
\label{expn}
n=\frac{x^2}{1-(m+1) x^2}=-\frac{1}{m_0}+\frac{m-m_0}{m_0^2}-\frac{a_0}{m_0^3} (m-m_0)^2+O((m-m_0)^3)\\
 \  \ \textrm{with} \quad\quad
a_0=1-\frac{\left(\kappa ^2 m_0+m_0-1\right) \left(\left(m_0-1\right)
  \mathds{ K}\left(m_0\right)+\mathds{E}\left(m_0\right)\right){}^2}{4 \kappa ^2 \left(m_0-1\right) m_0}.
  \end{split}
\ee

Next we can  use  eqs. \eqref{expx}  and  \eqref{expn} to evaluate the expansion of $g(\tilde\sigma)$ and $g'(\tilde\sigma)$, two  quantities that often appears in our analysis.  The simplest way to calculate  $g(\tilde\sigma)$  is to use eq. \eqref{eq:2.34} and eq. \eqref{eq:2.36} to obtain an expression as a function of $n, m, x$ and $\kappa$.  We find
\begin{align}
\label{gm0}
g(\tilde\sigma)=&\sqrt{-\frac{2\left(\kappa ^2+1\right) m n x^2}{\kappa ^2+\left(\kappa
   ^2+1\right) (-(m+1)) x^2+\sqrt{\left(\kappa ^2-\left(\kappa ^2+1\right) (m+1)
   x^2\right)^2-4 \left(\kappa ^2+1\right) m x^4}}}=\nonumber\\=&\frac{1}{\sqrt{-m_0}}+\frac{m-m_0}{2
   \left(-m_0\right){}^{3/2}}+\nonumber\\
   &+\frac{ (4 a_0-1)
   \left(m_0-1\right)-\left(\left(m_0-1\right)
   \mathds{K}\left(m_0\right)+\mathds{E}\left(m_0\right)\right){}^2}{8 \left(m_0-1\right)
   \left(-m_0\right){}^{5/2}}\left(m-m_0\right){}^2+O\left(\left(m-m_0\right){}^3\right).
\end{align}
Similarly for  $g'(\tilde\sigma)$ we have 
\begin{align}
\label{gpm0}
g'(\tilde\sigma)=&-\kappa\sqrt{-m n^2-g^2(\tilde\sigma)}=\frac{\left(m-m_0\right) \left(\left(m_0-1\right)
   \mathds{K}\left(m_0\right)+\mathds{E}\left(m_0\right)\right)}{2
   m_0^2}+O\left(\left(m-m_0\right){}^2\right).
\end{align}
If we now plug these results into eq. \eqref{eq:3.6} for the derivative of $\eta$, we get at $m=m_0$:
\be
\begin{split}
\left.\frac{\partial\eta}{\partial m}\right|_{m=m_0}
   =&\frac{\left(1-m_0\right) \left(\mathds{E}\left(m_0\right)+(m_0-1)
   \mathds{K}\left(m_0\right)\right){}^2+m_0\kappa^2\left(1-\left(\mathds{E}\left(m_0\right)+\left(m_0-1\right)
   \mathds{K}\left(m_0\right)\right){}^2\right)}{2 \kappa 
   \left(1-m_0\right) \left(-m_0\right){}^{3/2} \sqrt{1-\left(\kappa ^2+1\right) m_0}}.
   \end{split}
\ee
This value of the derivative has been instrumental in sec. \ref{subsect:3.1} to investigate the monotonicity of $\eta$ with $m$.
\section{The Expansion of then renormalized  area  for \texorpdfstring{$\kappa\to\infty$}{kappatoinfty}}
\label{app:largekespansion}
Our goal, here, is to expand the renormalized area  for $\kappa\to\infty$ while keeping constant the distance $L$ from the defect.
From the numerical analysis in sec. \ref{subsect:3.1}, we see that the same value of $L$ is reached for smaller and smaller value of 
$m$ as $\kappa$ approaches infinity (at least if $\chi\ne\frac{\pi}{2}$). Thus we shall assume 
\be
m\to 0 \ \ \ \ \ \ \  \mbox{when} \ \ \  \kappa\to\infty\ \ \  \mbox{but $L$ is fixed.}
\ee
In this limit we can easily check that   $x\to 1$ by solving eq. \eqref{eq:2.36}.  We are motivated therefore to postulate the following expansion of
$m$ and $x$  for large flux 
\be
\label{explargeflux1}
m=\sum_{n=1}^\infty \frac{a_{2n}}{\kappa^{2n}} \ \ \ \  \ \ \ \  x=1+\sum_{n=1}^\infty \frac{b_{2n}}{\kappa^{2n}}.
\ee
At this level it is just an ansatz that will be justified by its consistency. Expanding in this way eq.  \eqref{eq:2.36} and eq. \eqref{eq:3.1} is potentially a delicate issue since  all  the entries of the elliptic functions  depend explicitly or implicitly on the modulus $m$: the results  below are obtained  by first computing the elliptic function  for small modulus keeping the other entries  fixed  and subsequently expanding the dependence on the other entries for small $m$. 

With this procedure eq. \eqref{eq:2.36} determining the flux reads
\begin{align}
\kappa^2=&
\frac{\kappa ^2}{a_2 \cos ^2\chi +2 b_2 \cot ^2\chi }+\frac{1}{8 \left(a_2 \sin ^2\chi +2
   b_2\right){}^2}\left[\tan \chi  \left(-4 \tan \chi  \left(a_4 \sin ^2\chi +2 b_4\right)+\right.\right.\nonumber\\
&+2 a_2 b_2 \sec ^2\chi 
   \left((\pi -2 \chi ) \left(\sin ^22\chi +2 \cos 2\chi \right)-4 \sin 2\chi -\sin 4 \chi\right)-a_2^2 \sin \chi  \tan ^2\chi~ \times\\ &\left.\left.\times  (4 \chi  \sin \chi -2
   \pi  \sin \chi +7 \cos \chi +\cos 3 \chi )+4 b_2^2 (8 \chi +3 \sin 2\chi -4 \pi ) \sec ^2\chi \right)\right]+O\left(\kappa^{-2}\right)\nonumber\
\end{align}
and similarly eq. \eqref{eq:3.1}, which instead fixes the distance, takes the form
\begin{align}
\frac{L}{R}=&
\tanh ^{-1}\left(\frac{\sqrt{-\frac{\left(a_2+2 b_2\right) \tan ^2\chi }{b_2}}}{\sqrt{2}}\right)+\frac{1}{2 \sqrt{2} \kappa ^2}\Biggl[\left(-a_2-2 b_2\right){}^{1/2} \sqrt{b_2} (\pi-2 \chi -\sin 2\chi)-\nonumber\\ &-\frac{2 \left(2 a_2^2 b_2+a_2 \left(11 b_2^2-b_4\right)+2 b_2 \left(a_4+b_2^2+b_4\right)\right) \sin ^2\chi }{b_2
   \left(a_2+4 b_2-a_2\cos 2\chi \right) \sqrt{-\frac{\left(a_2+2 b_2\right) \tan ^2\chi }{b_2}}}+\\ &+\!\frac{ \left(a_2\!+\!2 b_2\right) \left(a_2 \sin
   4 \chi\! -\! 2 (\pi\! -\! 2 \chi ) \left(a_2\!-\!4 b_2\right)\right)\!+\!2 \left(4 a_2 b_2\!+\!2 a_2^2\!+\!a_4\!-\!6 b_2^2\!+\!2 b_4\right) \sin 2\chi }{2 \left(a_2+4
   b_2-a_2\cos 2\chi \right) \sqrt{-\frac{\left(a_2+2 b_2\right) }{b_2}}}\Biggr]\!+\!O\left(\kappa^{-4}\right)\nonumber
\end{align}
We can now solve iteratively  this combined system of equations and determine $\{a_2, a_4,b_2, b_4\}$ in terms of $L/R$ and $\chi$.
We get
\begin{align}
a_2=&\frac{1}{2} \sec ^4\chi\left(-\frac{2 L^2}{R^2}+\cos 2 \chi -1\right)\\
b_2=&\frac{1}{2} \left(\frac{L^2}{R^2}+1\right) \tan ^2\chi  \sec ^2\chi \\
a_4=&-\frac{1}{32} \sec ^9\chi  \left(-\frac{2 L^2}{R^2}+\cos 2\chi -1\right) \left(-\frac{88 L^2 \chi  \sin \chi }{R^2}+\frac{44 \pi  L^2 \sin \chi }{R^2}+\frac{8
   L^2 \chi  \sin 3\chi }{R^2}-\right.\nonumber\\ &-\left.\frac{4 \pi  L^2 \sin 3\chi }{R^2}-2 \left(\frac{16 L^2}{R^2}+5\right) \cos \chi -84 \chi  \sin \chi +42 \pi  \sin \chi +12
   \chi  \sin 3\chi -\right.\nonumber\\ &\left.\phantom{-\frac{2 L^2}{R^2}}\!\!\!\!\!\!\!\!\!\!\!\!\!\!\!\!\!-6 \pi  \sin 3\chi +9 \cos 3\chi +\cos 5 \chi \right)\\
b_4=&-\frac{1}{64} \left(\frac{L^2}{R^2}+1\right) \tan \chi  \sec ^8\chi  \left(-\frac{112 L^2 \chi }{R^2}-\frac{52 L^2 \sin 2\chi }{R^2}+\frac{2 L^2 \sin 4 \chi
   }{R^2}-\right.\nonumber\\ &\left.-8 (\pi -2 \chi ) \left(\frac{5 L^2}{R^2}+6\right) \cos 2\chi +\frac{56 \pi  L^2}{R^2}-84 \chi -31 \sin 2\chi +14 \sin 4\chi +\right.\nonumber\\ &\left.\phantom{-\frac{2 L^2}{R^2}}\!\!\!\!\!\!\!\!\!\!\!\!\!\!\!\!\!+\sin 6\chi +6 (\pi -2
   \chi ) \cos 4\chi +42 \pi \right).
\end{align}
We can also use eq. \eqref{explargeflux1} to expand the renormalized area in terms of the above parameters and, adopting the same prescription, one obtains 
\begin{align}
\frac{S}{\sqrt{\lambda}}=&
-\kappa  \sqrt{-\frac{1}{a_2+2 b_2}} \tan \chi -\frac{1}{8
   \kappa }\left(-\frac{1}{a_2+2 b_2}\right)^{3/2} \sec ^2\chi  \left[4 b_2 (\pi -2 \chi ) \left(a_2+2
   b_2\right)+\right. \nonumber\\ &+\left.\left(a_2^2+a_4-6 b_2^2\right) \sin 2\chi +2 \left(a_2 b_2+b_4\right) \sin2 \chi +a_2 (\pi -2 \chi ) \left(a_2+2 b_2\right) \cos 2\chi \right]+O\left({\kappa }^{-3/2}\right)
\end{align}
The explicit form of the coefficients $a_2,a_4,b_2$ and $b_4$ leads to the following expansion for the renormalized area
\be
\frac{S}{\sqrt{\lambda}}=
-\frac{\kappa  R \sin \chi }{L}-\frac{R (2 \chi +\sin 2\chi -\pi ) \sec ^3\chi  \left(-\frac{2 L^2}{R^2}+\cos 2\chi -1\right)}{16 \kappa 
   L}+O\left({\kappa }^{-2}\right).
\ee

\section{The \texorpdfstring{$\chi=\frac{\pi}{2}$}{chieqpidiv2}  Wilson loop}
\label{pidiv2}

\noindent
The value $\chi=\frac{\pi}{2}$ is peculiar since it corresponds to the absence of motion on the internal sphere $S^5$. In fact, in this case  $\theta$ already takes its  maximum value at the boundary ($\sigma=0$) and thus it cannot be increased further. From eq. \eqref{eq:2.11}, this results in setting $j=0$ and consequently  $m\leq -1$. Therefore we are in  the region (B) of the allowed parameters.  Since $j=0$, $\tilde \sigma$  becomes a free parameter and it can be determined by solving the equation for the flux
\be
\text{sn}\left(\left.\sqrt{-\frac{1}{m+1}} \sigma
   \right|m\right)^2={\frac{m+1}{2m}-\frac{\sqrt{\kappa ^2 (m+1)^2+(m-1)^2}}{2 m\sqrt{\kappa
   ^2+1}}},
\ee
and we get
\be
\label{D2}
\sqrt{n}\tilde \sigma=\sqrt{-\frac{1}{m+1}}\tilde\sigma=~
\text{sn}^{-1}\left. \left(\sqrt{\frac{m+1}{2m}-\frac{\sqrt{\kappa ^2 (m+1)^2+(m-1)^2}}{2 m\sqrt{\kappa
   ^2+1}}}
   \right|m\right).
\ee
When $m$ spans the entire interval from $-1$ to $-\infty$, $\tilde\sigma$ runs from $0$ to $\infty$. We can write $\eta$ in terms of 
$m$ 
\be
\eta=\tanh^{-1}\left(\sqrt{\frac{2 \kappa ^2}{1+\kappa ^2+\sqrt{\kappa ^2+1} \sqrt{\kappa ^2
  +\frac{(m-1)^2}{(1+m)^2}}}}\right)+\sqrt{\frac{m}{m+1}}\left[\sqrt{n}\tilde\sigma
-\Pi\left(\left.1+m; \text{am}(\sqrt{n}\tilde\sigma|m)\right|m\right)\right]
 \ee
\begin{wrapfigure}[13]{l}{90mm}
\centering 
\includegraphics[width=.45\textwidth]{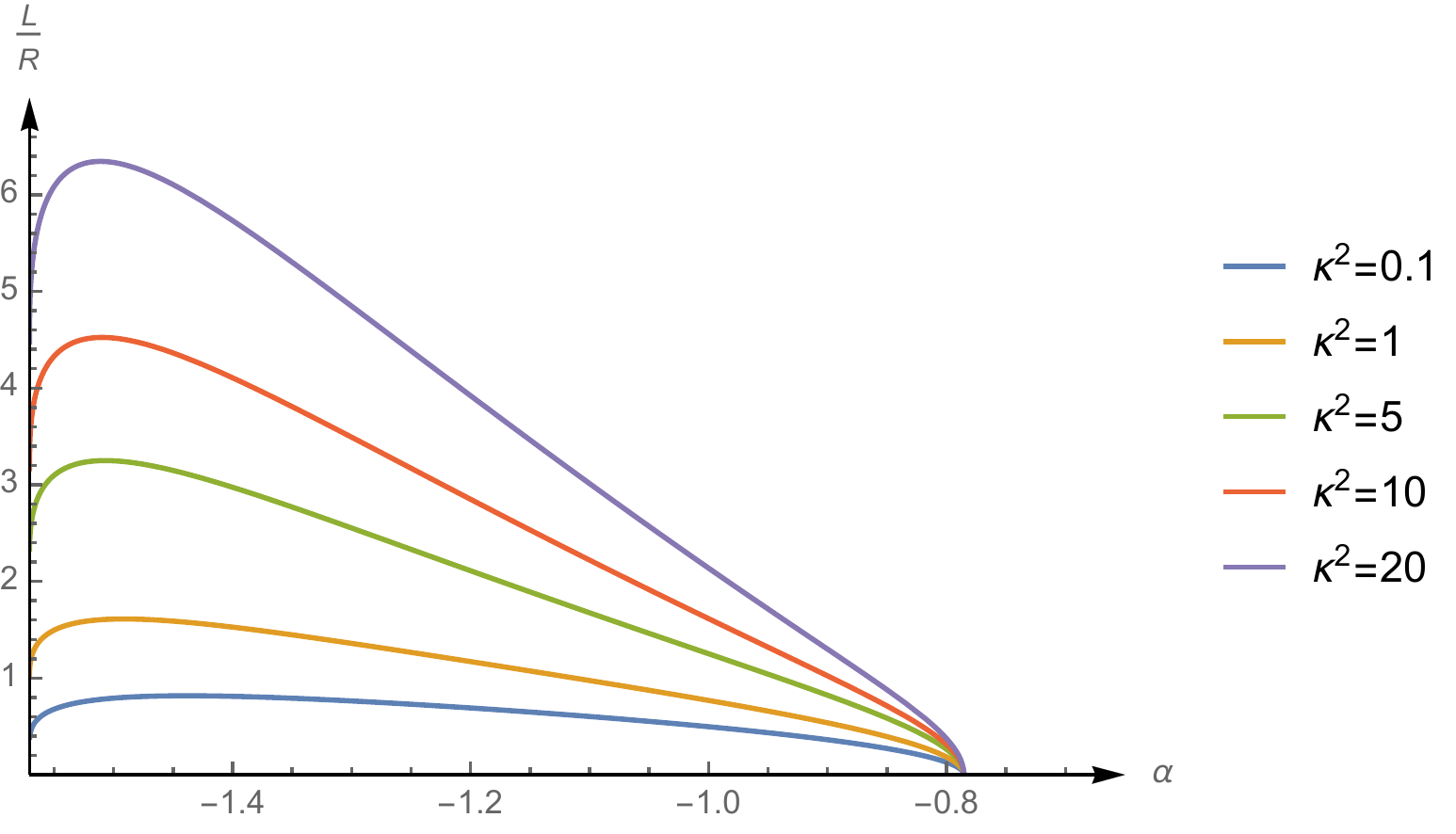}

\caption{\label{Distpi2} \footnotesize  The behavior of the distance from the defect as a function of $m$ at $\chi =\frac{\pi}{2}$ is not qualitatively different from the one obtained for other values of the angle greater than $\chi_s$. All the curves display a maximum.}
\end{wrapfigure} 
with the combination  $\sqrt{n}\tilde\sigma$ seen as a function of $m$ is given in eq. \eqref{D2}.   The behavior of the distance with $m$  is then  displayed  in 
fig.~\ref{Distpi2} where as usual we have parametrized  $m$ as $\tan\alpha.$ All the curves (independently of the value of the flux) are not monotonic  function of $m$ and display a maximum, which becomes steeper as the flux increases. The distance from the defect always vanishes  at $m=-1$.  This property can be checked analytically by means of the results of app.~\ref{distance}. Next we examine the area  given by eq. \eqref{eq:3.10}.  Its behavior is pictured in fig. \ref{Areapi2} and we again find that it is not a monotonic function of  $m$.   When  $\alpha=\arctan(m)$ increases, it reaches a maximum exactly for the same value of $\alpha$ for which the distance does and then it decreases to $-\infty$.  For $\alpha=-\frac{\pi}{2}$, namely $m\to -\infty$ all the curves go to the same value: in this case  $-1$.  Therefore we observe the same behavior previously obtained for all the angles $\chi>\chi_s$.

 \begin{figure}[h]\hskip 1cm
                      	\begin{minipage}{.45\textwidth}
	\includegraphics[width=.90\textwidth]{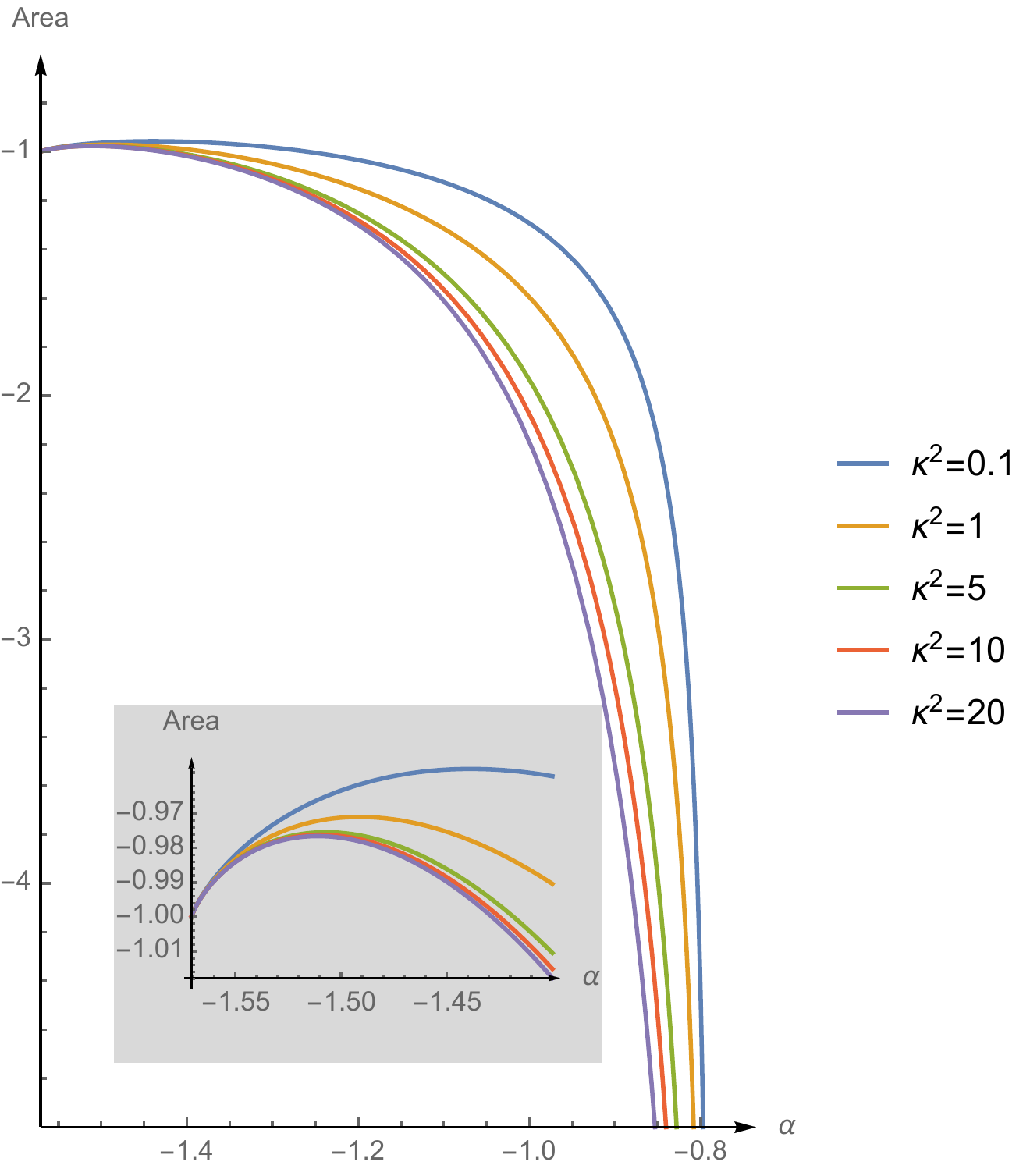}
	\caption{\label{Areapi2} \footnotesize  The behavior of the area of the connected surface  as a function of $m$ at $\chi =\frac{\pi}{2}$ is not qualitatively different from the other values of the angle greater than $\chi_s$. All the curves display a maximum}
	\end{minipage}\hskip 1cm
	                      	\begin{minipage}{.45\textwidth}\vskip 1.5cm
		\includegraphics[width=.90\textwidth]{{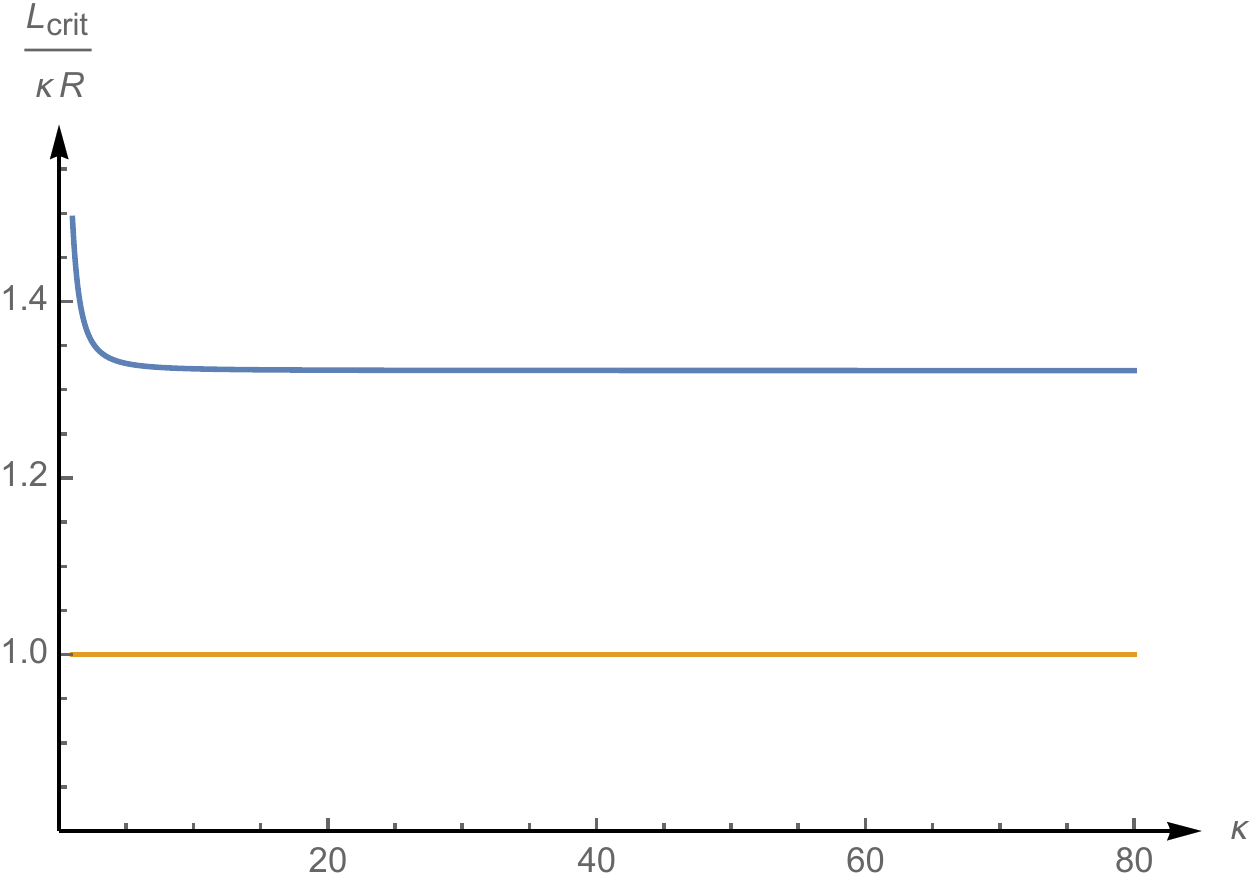}}
				\caption{\label{LsuKR} \footnotesize Plot of $\frac{L_{\rm cric.}}{\kappa R}$ as function of $\kappa$. This quantity is decreasing with the flux, but always greater than one.}
			\end{minipage}
\end{figure}

Finally, we  investigate the phase transition from the dome to the connected solution when we vary the distance from the defect.
For $\chi=\frac{\pi}{2}$  the disconnected solution cannot exist for all distances. In fact, when  $\frac{L}{\kappa R}<1$, the dome solution is no more acceptable because it intersects the (defect) brane. This phenomenus could provides a second putative mechanism for the phase transition from a disconnected to a connected minimal surface: even if the area of the latter might become dominant only when $\frac{L}{\kappa R}<1$,  we are forced to start using it at $\frac{L}{\kappa R}=1$. In this case the transition is of order zero.

However this  second  mechanism remains inoperative if the area of the connected solutions becomes smaller than $-1$ before the dome touches the brane.
To explore this point we have plotted in fig. \ref{LsuKR} the critical distance divided by $\kappa R$  for different values of the flux. This quantity is monotonically decreasing with $\kappa$, but it is always greater than one. Thus the first-order transition always occurs before the dome touches the brane.

\section{Connected solution as correlator between two circles of different radii}
\label{circles}
In this appendix we show that our extremal surface can be viewed as the solution connecting two coaxial circles of different radii and different couplings with the scalars. The former is identified with the original one.  The latter is located behind the defect, and the distance from it is chosen so  that the brane intersects the extremal surface orthogonally.

To begin with, we shall examine more carefully  the geometric structure of our solution. In $AdS_5$ our connected solution is given by  
\be
\begin{gathered}
\label{E1}
y(\sigma)=\frac{ R\cosh\eta}{\sqrt{1+g^2(\sigma)}} \textrm{sech} [v(\sigma)-\eta]\quad\quad
r(\sigma)=R\cosh\eta \frac{g(\sigma)}{\sqrt{1+g^2(\sigma)}} \textrm{sech} [v(\sigma)-\eta]\\
x_3(\sigma)= -R\cosh\eta\tanh [v(\sigma)-\eta],
\end{gathered}
\ee
and it is confined into a $S^3$ inside AdS$_5$. In fact
\be
\begin{split}
x^2+y^2+r^2=&R^2\cosh^2\eta\left[\frac{ \textrm{sech}^2 [v(\sigma)-\eta]}{1+g^2(\sigma)} +
 \frac{g^2(\sigma)}{1+g^2(\sigma)} \textrm{sech}^2 [v(\sigma)-\eta]+\tanh^2 [v(\sigma)-\eta]\right]=\\
 =&R^2\cosh^2\eta=\sqrt{L^2+R^2}
 \end{split}
\ee
The surface \eqref{E1} intersects the boundary of AdS$_5$ at $\sigma=0$ and if we extend the range of the world-sheet coordinate $\sigma$ beyond $\tilde \sigma$, it reaches again the boundary at    $\hat\sigma=\frac{1}{\sqrt{n}}\mathds{K}(m)$ for which
\be
v(\hat\sigma)=2\sqrt{\frac{\epsilon_0}{n}}\left[ \mathds{K}(m)-\Pi\left(-\frac{1}{n},m\right)\right].
\ee
This second intersection with the  boundary of AdS$_5$ is again a circle of  radius
\be
\hat{R}=R\cosh\eta~ \textrm{sech} \left(2\sqrt{\frac{\epsilon_0}{n}}\left[ \mathds{K}(m)-\Pi\left(-\frac{1}{n},m\right)\right]-\eta\right).
\ee
This second circle is located  at
\be
\begin{split}
x_3(\hat\sigma)=& -R\cosh\eta\tanh [v(\hat \sigma)-\eta]=R\sinh\eta-R\sinh v(\hat\sigma)\textrm{sech}\left( v(\hat\sigma)-\eta\right)=\\
=&
R\sinh\eta-\hat R\textrm{sech}\eta \sinh v(\hat\sigma)
\end{split}
\ee
and the distance $\ell$ in the transverse direction $x_3$ between this circle and the one at $\sigma=0$ is 
\be
\ell=\hat R\textrm{sech}\eta \sinh v(\hat\sigma).
\ee
The angle $\chi$ describing the scalars coupling to the two circles is different and the $\Delta\chi$ between the two loops is
\be
\Delta\chi=\theta(\hat\sigma)-\theta(0)=\frac{2 j}{\sqrt{n}} \mathds{K}(m).
\ee
\section{Basis for chiral primary operators}
\label{Basis}
As already pointed out in the original papers, the only operators built from scalars  that can have  one-point functions different from zero are those invariant under $SO(3)\times SO(3)$.
If we consider the case of operator with classical dimension $2$ we can construct only two operators of this type: the famous Konishi operator
\be
\mathcal{K}(x)=\frac{4\p^2}{\sqrt{3}}\mathrm{Tr}(\Phi_I \Phi^I)
\ee
and the chiral primary operator 
\be
\mathfrak{O}_1(x)=\sqrt{\frac{1}{6}}\mathrm{Tr}\left[\Phi_1^2+\Phi_2^2+\Phi_3^2-\Phi_4^2-\Phi_5^2-\Phi_6^2\right].
\ee
The list of CPO's is completed by $4$ diagonal operators
\be
\begin{split}
\mathfrak{O}_2(x)=\sqrt{\frac{1}{2}}\mathrm{Tr}\left[\Phi_1^2-\Phi_3^2\right]\ \ \ \ \ \ \ \mathfrak{O}_3(x)=\sqrt{\frac{1}{6}}\mathrm{Tr}\left[\Phi_1^2-2\Phi_2^2+\Phi_3^2\right]\\
\mathfrak{O}_4(x)=\sqrt{\frac{1}{2}}\mathrm{Tr}\left[\Phi_4^2-\Phi_6^2\right]\ \ \ \ \ \ \ \mathfrak{O}_5(x)=\sqrt{\frac{1}{6}}\mathrm{Tr}\left[\Phi_4^2-2\Phi_5^2+\Phi_6^2\right]
\end{split}
\ee
and $15$ off-diagonal ones $\mathfrak{O}_{IJ}(x)=\sqrt{2}\mathrm{Tr}\left[\Phi_I\Phi_J\right]$ with $i< j$. When performing integrability calculation, one usually computes the
expectation values of the CPO $\mathrm{Tr}(Z^2)$. We want to expand this operator in our basis and we find
\be
\begin{split}
\mathrm{Tr}(Z^2)=&\mathrm{Tr}(\Phi_3^2-\Phi_6^2) +2 i \mathrm{Tr}(\Phi_3\Phi_6) =\\
=&\sqrt{\frac{2}{3}} \mathfrak{O}_1(x)-\sqrt{\frac{1}{2}} \mathfrak{O}_2(x)+\sqrt{\frac{1}{6}} \mathfrak{O}_3(x)+\sqrt{\frac{1}{2}} 
\mathfrak{O}_4(x)-\sqrt{\frac{1}{6}} \mathfrak{O}_5(x)+\sqrt{2} i  \mathfrak{O}_{36}(x) .
\end{split}
\ee
Then we have the following relation between the VEV's:
\be
\langle \mathfrak{O}_1(x) \rangle_0=\sqrt{\frac{3}{2}}\langle\mathrm{Tr}(Z^2) \rangle_0.
\ee
Actually in the Wilson loop operator appears the following linear combination of CPO $ Y^a(\theta)\mathcal{O}_a(x) $, where $Y^a(\theta)=(C^a_{IJ}\theta^I\theta^J)=\left(-\sqrt{\frac{1}{6}},0,0,-\sqrt{\frac{1}{2}},\sqrt{\frac{1}{6}},\mathbf{0}\right)$ is a $20$ component vector. The boldface zero indicates that the remaining $15$ components vanishes
and the $C^a_{IJ}$ form the   basis of the symmetric traceless tensors  that we used for constructing the CPO. Then
\be
\langle Y^a(\theta)\mathcal{O}_a(x) \rangle=\langle Y^1(\theta)\mathcal{O}_1(x) \rangle=4\sqrt{2}\pi^2\langle Y^1(\theta)\mathfrak{O}_1(x) \rangle=
-\frac{4\pi^2}{\sqrt{2}}\langle\mathrm{Tr}(Z^2) \rangle
\ee
 Now if we use eq. (28) of \cite{Buhl-Mortensen:2017ind}, we find up to one-loop
\be
\begin{split}
\langle\mathrm{Tr}(Z^2) \rangle=&-\frac{1}{12 L^2} k(1-k^2)\!+\! \frac{g^2_{YM}}{8\pi^2L^2} \left(k(N-k)\!+\!\frac{(k-1)}{2}\!+\!\sum^{\left[\frac{k-2}{2}\right]}_{i=0}(H_{k-i-1}-H_i)(k-2i -1)\right)=\\
=&-\frac{1}{12 L^2} k(1-k^2)+ \frac{g^2_{YM}}{8\pi^2L^2} \left(k(N-k)+\frac{(k-1)}{2}+\frac{k(k-1)}{2}\right)=\\
=&-\frac{1}{12 L^2} k(1-k^2)+ \frac{g^2_{YM}}{8\pi^2L^2} \left(k(N-k)+\frac{k^2-1}{2}\right)
\end{split}
\ee
Therefore 
\be
\langle Y^a(\theta)\mathcal{O}_a(x) \rangle=-\frac{4\pi^2}{\sqrt{2}}\langle\mathrm{Tr}(Z^2) \rangle=
\frac{\pi^2}{3\sqrt{2} L^2} k(1-k^2)- \frac{g^2_{YM}}{2\sqrt{2}L^2} \left(k(N-k)+\frac{k^2-1}{2}\right)
\ee
For the Konishi operator up to one-loop instead we have \cite{Buhl-Mortensen:2017ind}
\be
\langle\mathcal{K}(x)\rangle=-\frac{\pi^2}{\sqrt{3} L^2} k(1-k^2)-\frac{\sqrt{3}\lambda }{4 L^2}   k\left(1-k^2\right) \left(\psi\left(\frac{k+1}{2}\right)+\gamma_E-\log 2+\frac{5}{6}\right)
\ee
\bibliographystyle{JHEP}
\bibliography{biblioDCFT} 

\end{document}